\PassOptionsToPackage{pdfpagelabels=false}{hyperref} 
\documentclass[letter]{aa}

\usepackage{natbib}
\usepackage{graphicx}
\usepackage{txfonts}
\usepackage{amsmath}
\usepackage{amssymb}	
\usepackage{rotating}
\usepackage{natbib}
\usepackage{gensymb}

\bibpunct{(}{)}{;}{a}{}{,} 
\usepackage[colorlinks,linkcolor=blue,anchorcolor=blue,citecolor=blue]{hyperref}

\begin{document}


\title{Ultra-low frequency LOFAR spectral indices of cluster radio halos}

\author{T. Pasini\inst{\ref{inst1},\ref{inst2}}
\and F. De Gasperin\inst{\ref{inst1}}
\and M. Br\"uggen\inst{\ref{inst2}}
\and R. Cassano\inst{\ref{inst1}}
\and A. Botteon\inst{\ref{inst1}}
\and G. Brunetti\inst{\ref{inst1}}
\and H. W. Edler\inst{\ref{inst2}}
\and R. J. van Weeren\inst{\ref{inst3}}
\and V. Cuciti\inst{\ref{inst1}, \ref{inst2}, \ref{inst4}}
\and T. Shimwell\inst{\ref{inst3}, \ref{inst5}}
\and G. Di Gennaro\inst{\ref{inst2}}
\and M. Gaspari\inst{\ref{inst6}}
\and M. Hardcastle\inst{\ref{inst7}}
\and H. J. A. Rottgering\inst{\ref{inst3}}
\and C. Tasse\inst{\ref{inst8}, \ref{inst9}}
}
\authorrunning{T. Pasini, F. De Gasperin, M. Br\"uggen et al.}

\institute{INAF - Istituto di Radioastronomia, via P. Gobetti 101, 40129, Bologna, Italy\label{inst1}\\
\email{thomas.pasini@inaf.it}
\and Hamburger Sternwarte, Universität Hamburg, Gojenbergsweg 112, 21029 Hamburg, Germany\label{inst2}
\and Leiden Observatory, Leiden University, PO Box 9513, 2300 RA Leiden, The Netherlands\label{inst3}
\and Dipartimento di Fisica e Astronomia, Università di Bologna, Via P. Gobetti 93/2, 40129 Bologna, Italy\label{inst4}
\and ASTRON, the Netherlands Institute for Radio Astronomy, Oude Hoogeveensedijk 4, 7991 PD Dwingeloo, The Netherlands\label{inst5}
\and Department of Astrophysical Sciences, Princeton University, Princeton, NJ 08544, USA\label{inst6}
\and Department of Physics, Astronomy and Mathematics, University of Hertfordshire, College Lane, Hatfield AL10 9AB, UK\label{inst7}
\and GEPI \& ORN, Observatoire de Paris, Université PSL, CNRS, 5 Place Jules Janssen, 92190 Meudon, France\label{inst8}
\and Department of Physics \& Electronics, Rhodes University, PO Box 94, Grahamstown, 6140, South Africa\label{inst9}
}



\abstract
{A fraction of galaxy clusters harbor diffuse radio sources known as radio halos. The currently adopted scenario for their formation is based on second-order Fermi re-acceleration of seed electrons that is driven by merger-driven turbulence in the intra-cluster medium. This mechanism is expected to be inefficient, which implies that a significant fraction of halos should have very steep energy spectra ($\alpha < -1.5$).}
{We start investigating the potential and current limitations of the combination of the two surveys conducted by LOFAR, LoTSS (144 MHz), and LoLSS (54 MHz), to probe the origin of radio halos.}
{We follow up the 20 radio halos detected in the DR1 of LoTSS, which covers the HETDEX field, with the LoLSS survey, and we study their spectral properties between 54 and 144 MHz.}
{After the removal of compact sources, 9 halos were excluded due to unreliable halo flux density measurements at 54 MHz. Our main finding is that 7 out of 11 ($\sim$64\%) exhibit an ultra-steep spectrum ($\alpha < -1.5$), which is a key prediction of turbulent re-acceleration models. We also note a tentative trend for more massive systems to host flatter halos, although the currently poor statistics does not allow for a deeper analysis.}
{Our sample suffers from low angular resolution at 54 MHz, which limits the accuracy of the compact-sources subtraction. Nevertheless, this study is the first step towards providing compelling evidence for the existence of a large fraction of radio halos with very steep spectrum, which is a fundamental prediction of turbulent re-acceleration models. In this regard, the forthcoming second data release of LoLSS, along with the integration of LOFAR international stations and the instrumental upgrade to LOFAR2.0, will improve both the statistics and the low-frequency angular resolution, allowing to conclusively determine the origin of radio halos in galaxy clusters.}


\maketitle


\section{Introduction} 
\label{sec:intro}

Radio halos are diffuse sources observed in the central regions of disturbed galaxy clusters \citep{Willson_1970, vanWeeren_2019}. They trace synchrotron emission by relativistic Cosmic-Ray (CR) electrons in the presence of magnetic fields in the Intra-Cluster Medium (ICM, \citealt{Bruggen_2012, Brunetti_2014}), the hot ($\sim 10^7 - 10^8$ K) plasma which permeates galaxy clusters. The synchrotron spectrum of radio halos can be described to a first approximation by a power-law $S_\nu \propto \nu^{\alpha}$, with $S_\nu$ the flux density at a frequency $\nu$, and $\alpha$ the spectral index. Given their (usually) $\sim$Mpc size, the electrons producing the radio emission need to be re-energised or injected \textit{in-situ} since they cannot fill the emitting volume within their synchrotron cooling time \citep{Jaffe_1977}. Their origin has been attributed to either re-acceleration by merger-driven subsonic turbulence (turbulent models, \citealt{Brunetti_2001, Petrosian_2001, Cassano_2005, Brunetti_2007, Beresnyak_2013, Donnert_2014, Miniati_2015}) or injection of secondary electrons \textit{in-situ} by proton-proton collisions (hadronic models, \citealt{Dolag_2000, Pfrommer_2008, Ensslin_2011}). The observed correlation between radio halos and the cluster dynamical status, with halos being usually associated with more dynamically-disturbed systems, rather than with more relaxed clusters \citep{Buote_2001, Cassano_2010a, Cuciti_2021, Cassano_2023}, together with gamma-ray constraints \citep{Brunetti_2017, Adam_2021} and arguments based on the CR protons (CRp) energy budget in the case of steep-spectrum radio halos \citep{Brunetti_2008, Brunetti_2014, Bruno_2021}, support the turbulent re-acceleration scenario. Therefore purely hadronic models are currently disfavoured, but they might still play a role \citep{Brunetti_2011, Pinzke_2017, Adam_2021, Nishiwaki_2021}.

Turbulent re-acceleration models assume that mergers between galaxy clusters can generate turbulence in the ICM, which amplifies seed magnetic fields \citep{Dolag_2005} and re-accelerates relativistic particles via second-order Fermi mechanisms \citep{Brunetti_2001, Cassano_2005, Brunetti_2007, Pinzke_2017, Nishiwaki_2021}. However, due to the low efficiency of this mechanism (see e.g. \citealt{Brunetti_2011}), these models predict the existence of a large population of radio halos with very steep spectra \citep{Cassano_2006, Brunetti_2008}. Specifically, due to synchrotron and Inverse Compton (IC) losses, the maximum energy of electrons that are re-accelerated by second-order mechanisms in the ICM is generally estimated to be in the range 1-10 GeV, implying a maximum synchrotron frequency ($\nu_{\rm b}$, \textit{break} frequency) in the radio band and a gradual steepening of the synchrotron spectrum above this frequency \citep{Cassano_2010b}, which is proportional to the ICM acceleration efficiency $\chi$ \citep{Cassano_2006}:

\begin{equation}
\centering
\nu_b \propto \dfrac{B \chi^2}{(B^2 + B^2_{\rm CMB})^2} \ ,
\label{eq:efficiency}
\end{equation}

\noindent
where $B$ represents the cluster magnetic field, while $B_{\rm CMB} = 3.2 (1+z)^2\,\mathrm{\mu{G}}$ is the equivalent magnetic field strength of the Cosmic Microwave Background (CMB) at redshift \textit{z}. In this scenario, the acceleration efficiency depends on the specific acceleration mechanism and turbulent properties of the ICM. It is generally expected to increase with cluster mass, as the turbulent energy budget and turbulent energy flux in more massive clusters are larger (e.g., $\chi \propto M^{4/3}$, with $M$ mass of the main cluster undergoing the merger\footnote{This comes from the proportionality between break frequency and mass, see e.g. \citet{Cassano_2005}.}, see also Eq. 3 of \citealt{Cassano_2010b}).
This leads to the prediction of turbulent models that less energetic merger events, i.e. minor mergers or mergers in less massive systems, generate radio halos with steeper spectra \citep{Cassano_2006, Brunetti_2007, Brunetti_2008}. Since the vast majority of mergers in the universe involve low-mass systems, these models predict the existence of a vast population of radio halos with very steep spectra \citep{Cassano_2010b}. These radio halos, referred to as ultra-steep spectrum radio halos (USSRH), are predicted to have synchrotron spectral indices $\alpha < -1.5$. This is steeper than the typical $\alpha \sim -1.2$---$-1.3$ found in radio halos observed in high-frequency radio surveys \citep[e.g.][]{Giovannini_1999}, that are usually associated with more massive clusters.

Thus, observations of USSRH constitute a unique tool to constrain the origin of radio halos. In recent years, a growing number of USSRH have been found through single-target studies of galaxy clusters \citep[e.g.][]{Brunetti_2008, Bonafede_2012, Wilber_2018, diGennaro_2021, Bruno_2021, Edler_2022}. The existence of these systems has also been used to further challenge an hadronic origin of giant radio halos \citep{Brunetti_2008, Bruno_2021}. However, it is still unclear whether they are hints of the emergence of a large population or if they constitute peculiar cases.  

In this letter, we explore the potential of the two radio surveys that are being carried out by the LOw Frequency ARray (LOFAR, \citealt{vanHaarlem_2013}) for the study of the spectral properties of radio halos in galaxy clusters. We will present the results based on the currently available data, demonstrating their potential impact but also their present weaknesses. Our results will also be put into context with state-of-the-art theoretical models for the origin of radio halos, with the caveat that the present data do not allow to derive any final conclusions on their nature. Throughout this work, we adopt a $\Lambda$CDM cosmology with H$_0 = 70$ km s$^{-1}$ Mpc$^{-1}$, $\Omega_\Lambda = 0.7$ and $\Omega_{\text{M}} =  1-\Omega_\Lambda  = 0.3$.

\section{Sample and spectral index measurement}
\label{sec:sample}

\begin{table*}[t!]
\centering
\begin{tabular}{l l l l l l l l}
\hline
\hline
Cluster name & S$_{1r_\mathrm{e}}^{54 \rm MHz}$ & S$_{3\sigma}^{54 \rm MHz}$ & S$_{1r_\mathrm{e}}^{144 \rm MHz}$ & S$_{3\sigma}^{144 \rm MHz}$ & $\alpha_{1r_\mathrm{e}}$ & $\alpha_{3\sigma}$ & Image \\
& [mJy] & [mJy] & [mJy] & [mJy] & & \\
\hline
PSZ2G086.93+53.18 & 6 $\pm$ 3 & 25 $\pm$ 7 & 0.7 $\pm$ 0.2 & 3.1 $\pm$ 0.5 & -2.20$\pm$ 0.61 & -2.15 $\pm$ 0.32 & T90 (=150 kpc) \\
PSZ2G096.83+52.49 & 122 $\pm$ 21 & 288 $\pm$ 43 & 20.8 $\pm$ 2.2 & 43.8 $\pm$ 4.5 & -1.80 $\pm$ 0.30 & -1.92 $\pm$ 0.17 & T100 kpc \\
PSZ2G099.86+58.45 & 34 $\pm$ 5 & 22 $\pm$ 4 & 6.5 $\pm$ 0.8 & 4.3 $\pm$ 0.6 & -1.67 $\pm$ 0.26 & -1.66 $\pm$ 0.27 & T50 kpc \\
PSZ2G107.10+65.32N & 205 $\pm$ 14 & 1084 $\pm$ 69 & 65.5 $\pm$ 6.6 & 314.8 $\pm$ 31.5 & -1.16 $\pm$ 0.24 & -1.26 $\pm$ 0.04 & T100 kpc \\
PSZ2G107.10+65.32S & 78 $\pm$ 15 & 248 $\pm$ 33 & 24.8 $\pm$ 2.9 & 83.7 $\pm$ 8.9 & -1.17 $\pm$ 0.46 & -1.11 $\pm$ 0.26 & T50 kpc \\
PSZ2G111.75+70.37 & 37 $\pm$ 7 & 205 $\pm$ 26 & 5.1 $\pm$ 0.6 & 27 $\pm$ 3.0 & -2.02 $\pm$ 0.27 & -2.07 $\pm$ 0.14 & T100 kpc \\
PSZ2G114.31+64.89 & 149 $\pm$ 12 & 324 $\pm$ 23 & 23.9 $\pm$ 2.4 & 43.2 $\pm$ 4.4 & -1.87 $\pm$ 0.16 & -2.05 $\pm$ 0.05 & T50 kpc \\
PSZ2G133.60+69.04 & 318 $\pm$ 33 & 590 $\pm$ 54 & 62.7 $\pm$ 6.3 & 115.8 $\pm$ 12.1 & -1.66 $\pm$ 0.20 & -1.66 $\pm$ 0.11 & T50 kpc\\
PSZ2G135.17+65.43 & 36 $\pm$ 8 & 50 $\pm$ 10 & 10.7 $\pm$ 1.2 & 16.6 $\pm$ 1.8 & -1.24 $\pm$ 0.47 & -1.12 $\pm$ 0.39 & T50 kpc\\
PSZ2G139.18+56.37 & 303 $\pm$ 20 & 1249 $\pm$ 82 & 96.5 $\pm$ 9.7 & 292.3 $\pm$ 29.3 & -1.17 $\pm$ 0.24 & -1.48 $\pm$ 0.04 & T50 kpc \\
PSZ2G143.26+65.24 & 33 $\pm$ 6 & 115 $\pm$ 14 & 7.3 $\pm$ 1.0 & 28.3 $\pm$ 3.0 & -1.54 $\pm$ 0.35 & -1.43 $\pm$ 0.19 & T100 kpc \\
\hline
\hline
\end{tabular}
\caption{\footnotesize The table lists, from first to last column: cluster name, 54 MHz halo flux density estimated within 1 $e-$folding radius and within 3$\sigma$ 144 MHz contours, 144 MHz halo flux density estimated within 1 $e$-folding radius and within 3$\sigma$ 144 MHz contours, spectral index estimated within 1 $e$-folding radius and within 3$\sigma$ 144 MHz contours, source-subtracted image used for flux density measurement (T50 kpc = taper 50 kpc, T100 kpc = taper 100 kpc, T90 = taper 90$''$).} 
\label{tab:sample}
\end{table*}

We use the Data Release 1 (DR1) of the LOFAR LBA Sky Survey (LoLSS, \citealt{deGasperin_2021, deGasperin_2023}), performed at a central frequency of 54 MHz, together with the DR2 of the LOFAR Two-Metre Sky Survey (LoTSS, \citealt{Shimwell_2017, Shimwell_2019, Shimwell_2022}) at 144 MHz. We examine emission from galaxy clusters in the HETDEX Spring field\footnote{RA: 11h to 16h and Dec: 45$\degree$ to 62$\degree$, \citet{Hill_2008}}, which has been mapped by both surveys for a total area of $650\,\rm{deg}^2$. Our aim is to measure the spectral index of radio halos in an as-large-as-possible sample of systems, estimate the fraction of USSRH, and thus constrain theoretical models. We start from the all-sky Planck Sunyaev-Zeldovich 2 (PSZ2) catalog of galaxy clusters \citep{Planck_2016}, which provides SZ masses, and match them to LoLSS DR1. If a system is covered by at least one pointing and the primary beam response at its position is above 30\%, 144 MHz LoTSS observations are also retrieved and the cluster is included in our initial sample. This sample contains 49 PSZ2 galaxy clusters with observations at both 54 and 144 MHz. For 33 of these systems, we also retrieved archival X-ray observations with either \textit{Chandra} or XMM-Newton \citep{Botteon_2022, Zhang_2023}. The calibration strategy for LOFAR data is described in Appendix \ref{app:data}, while a discussion of possible selection effects and biases is reported in Appendix \ref{app:selection}. After calibration, for all galaxy clusters in our sample we produced 54 and 144 MHz images at multiple angular resolutions. To avoid spurious emission from Active Galactic Nuclei (AGN), we subtracted compact sources from the \textit{uv-}data by removing all sources in the field of the target with a Largest Linear Size (LLS) smaller than 250\,kpc at each cluster redshift. Additional detail can be found in \ref{app:data}.

We then carefully inspected the 54 MHz images of our initial sample, and compared them with the 144 MHz LoTSS catalogs of radio halos by \citet{Botteon_2022} and \citet{vanWeeren_2021}, where a first classification of diffuse emission had already been performed. Among the 49 clusters of our sample, in these studies 19 were reported to host either a confirmed or a candidate\footnote{i.e. radio emission detected but no X-ray observation available, as defined in \citet{Botteon_2022}.} radio halo, with one system (PSZ2G107.10+65.32, i.e. A1758) previously known to host two \citep{Botteon_2018}. At 54 MHz, we were able to confirm the presence of diffuse emission even at low frequency in all these 19 clusters (20 halos in total). X-ray observations were eventually exploited to confirm the detection, when available. We note that we do not detect diffuse emission at 54 MHz in clusters for which no radio halos had previously been found at 144 MHz.

Since our aim is to include only radio halos whose classification and spectral index can be accurately determined, we decided to exclude a number of systems whose diffuse emission was too complicated to disentangle from spurious sources. This is mostly due to complex compact-source subtraction, uncertain classification at 54 MHz, and contamination by AGN emission. Furthermore, most of these systems did not have a conclusive classification at 144 MHz in \citet{vanWeeren_2021} and \citet{Botteon_2022}, and were labeled as candidate halos\footnote{Although this was purely based on the lack of X-ray observations able to confirm their nature.}. The masses of the excluded systems cover the entire mass range of our initial sample. Additionally, our selection criteria are independent of cluster mass and are solely dictated by instrumental limitations. 
In addition, we also applied a cut in redshift at $z <$ 0.7. This is motivated by the fact that, at higher redshift, a $\sim$250 kpc diffuse source (which is the threshold of our compact-source subtraction) would be smaller than $\sim$30$''$, which is roughly 1/3 of the lowest image beam used to detect our radio halos. Therefore, at this redshift it would not be possible to distinguish diffuse emission from spurious sources (e.g. remnant AGN plasma, leftovers of source subtraction etc.). Because of this threshold, one system (PSZ2G084.10+58.72, $z = 0.731$) was excluded from the final sample. In Table \ref{tab:fulllist} (see Appendix \ref{app:table}) we provide a complete list of all the initial 19 systems hosting diffuse emission, specifying whether they are included or excluded in the final sample and, in the latter case, we shortly explain the main reason. These limitations, that significantly affect our analysis, have a much higher impact in LoLSS compared to LoTSS, where the higher resolution allow for a more accurate subtraction of spurious emission. 

After this cut, the final sample includes 10 galaxy clusters with halos, including A1758 which hosts two, for a total of 11 radio halos. The complete list can be found in Table \ref{tab:sample}. We have then measured the flux density of our radio halos at 54 and 144 MHz, and calculated their integrated spectral index. The measurement is done from circular regions centered on the peak of the halo as detected at 144 MHz in \citet{Botteon_2022}, and with a radius corresponding to one e-folding radius ($r_\mathrm{e}$) as estimated at 144 MHz. We also tested the accuracy of this method by performing an additional flux density estimate at both frequency within 3$\sigma$ contours derived at 144 MHz. A thorough and detailed discussion of all these measurements, of their motivations and impact on our results can be found in Appendix \ref{app:calculations}.
The \textit{k}-corrected flux densities of our halos derived within $r_\mathrm{e}$ and within 3$\sigma$ contours and the resulting spectral indices\footnote{see Appendix \ref{app:calculations} for more details} are listed in Table \ref{tab:sample}. 

\section{Results and discussion}
\label{sec:results}

\begin{figure}[t!]
\centering
{\includegraphics[scale=0.42]{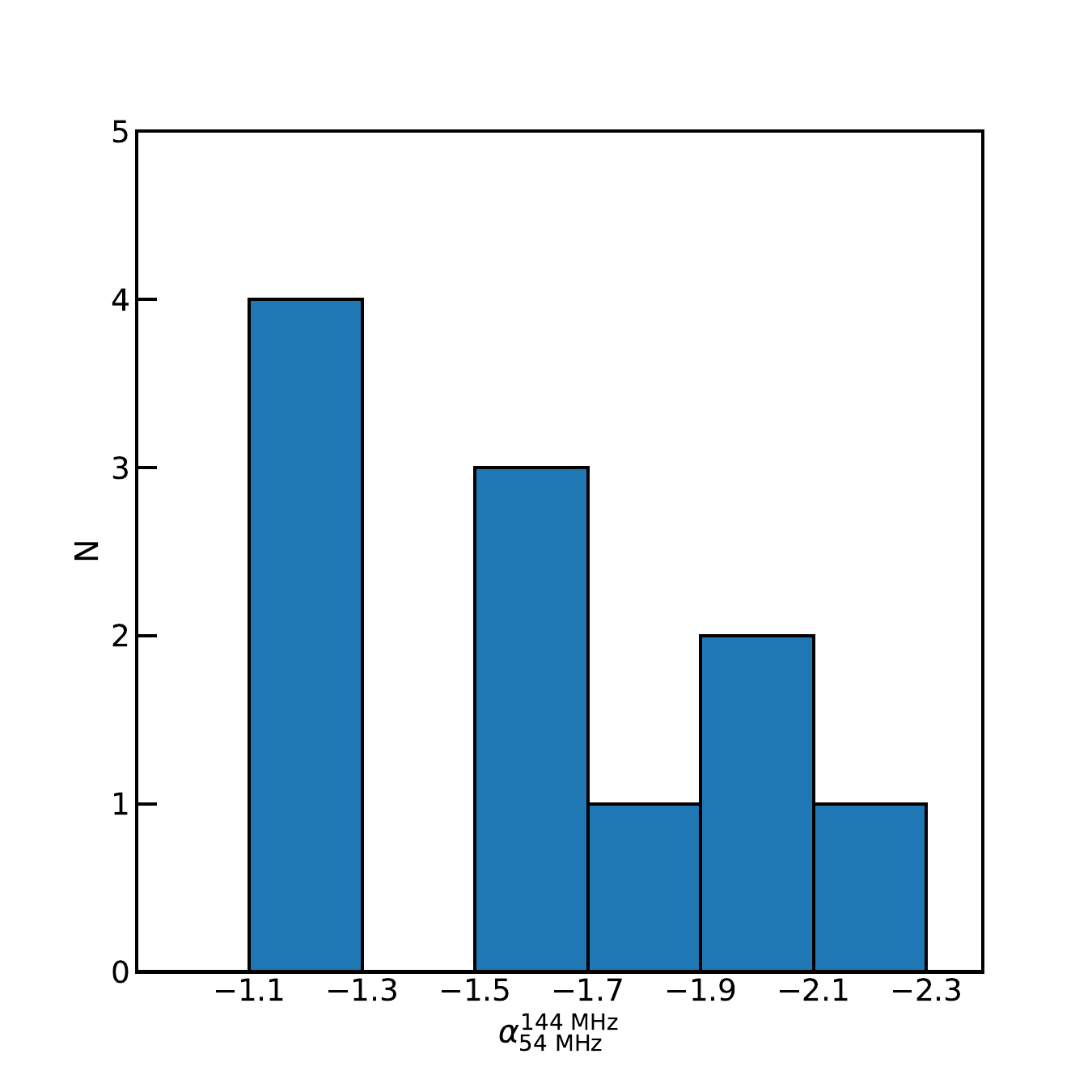}}
\caption{\footnotesize The spectral index distribution of our sample of 11 radio halos, estimated within a region centered on the halo and with radius 1~$r_\mathrm{e}$.}
\label{fig:hist}
\end{figure}

In Fig. \ref{fig:hist} we show the spectral index distribution for our sample. Out of 11 halos, seven ($\sim$64\%) exhibit $\alpha < -1.5$. Therefore, with the currently available data and techniques, a large fraction of the low-frequency ($<150$ MHz) radio halo population seems to show an ultra-steep spectrum, in line with predictions from e.g. \citet{Cassano_2010b}, who argued that more than half of radio halos that would have been observed with LOFAR would have a ultra-steep spectral index. This key prediction is tied to the fact that the spectra of radio halos are expected to become statistically steeper in radio halos generated in less massive systems \citep{Cassano_2006, Cuciti_2021}. Sunyaev-Zeldovich (SZ) mass estimates are available for our entire sample: the vast majority of these systems ($>80$\%) has $M < 7 \times 10^{14} M_\odot$, suggesting that a large fraction should host steep spectrum halos, which is consistent with our findings. While previous single-target studies (see Sec. \ref{sec:intro}) had already found a number of halos with steep spectra, this is the first time that this can be confirmed through a larger (albeit still small), self-consistent sample. Our result hints at the fact that previous $\sim$GHz surveys might have only scratched the surface of the halo population, since they were able to detect only halos with flat-enough spectra ($\alpha \sim -1.2$) to shine even at relatively high frequency. However, considering these findings, the vast majority of the halo population could instead exhibit ultra-steep spectra, which leads to very strong constrains on their origin. As thoroughly discussed in \citet{Brunetti_2008}, hadronic models struggle to explain the formation of such sources, since this would imply a steep spectral energy distribution for protons and, as a consequence, a domination of non-thermal protons in clusters. On the other hand, the existence of USSRH is naturally predicted from turbulent models, as already discussed in Sec. \ref{sec:intro}. This being said, it is important to note the limitations and potential sources of errors of our method. 

First of all, the starting sample of 20 radio halos has been significantly reduced to a point where the statistics, despite still being relevant, are too poor to derive any kind of conclusive result. With currently available data there is no way to improve it, apart from including systems that we previously excluded because of uncertain spectral index measurement, which would in any case introduce different kinds of uncertainties in our results. In the very next future LoLSS DR2 will, however, allow to increase the number of radio halos: \citet{Botteon_2022} finds a total of 80 galaxy clusters hosting radio halos in the DR2 of LoTSS. It is likely that, when combining it with LoLSS, we will still be forced to exclude a number of systems because of the same issues that impacted this work (low resolution, poor complex source subtraction etc.). If we assume to lose a similar fraction (around 50\%) of clusters, 40 radio halos will still constitute a significantly large sample, and the results will be considerably more robust. Finally, high-resolution observations that will be performed thanks to LOFAR International Stations (IS) and LOFAR2.0 in the next years will result in a much more accurate compact-sources subtraction.

Finally, it is worth noting that a potential tendency is detected for less massive systems in our sample to host halos with steeper spectrum, while massive clusters seem to exhibit flatter halos. While this is a crucial prediction of turbulent models, the statistics is too low to allow for any kind of deeper analysis. Thus, we reserve this discussion to a future work, where we will exploit the full DR2 of LoLSS to significantly increase the sample of halos. This is essential since, if confirmed, this would be a robust proof that radio halos with very steep spectrum are common in radio surveys at low frequencies. The results would clearly support the currently prevalent theoretical picture, which assumes that relativistic electrons in the ICM originate from re-acceleration mechanisms activated by cluster merger-driven turbulence, rather than from hadronic processes. Finally, it would conclusively indicate that a fraction of the kinetic energy associated with the motions of the matter on large scales is channelled into electromagnetic fluctuations in the plasma and on non-thermal components.

\section{Conclusions}

In this paper, we have combined data from currently available observations of the two LOFAR surveys, LoTSS and LoLSS, to derive integrated spectral indices for a sample of radio halos. We have discussed our results, including the potential sources of errors and limitations, and their impact on formation models of halos. Our results can be summarised as follows:

\begin{itemize}

\item We have followed-up with LoLSS a sample of 20 radio halos that were first observed with LoTSS. All these sources are also detected at 54 MHz. After compact-sources subtraction, we excluded 9 halos because of leftover AGN/compact emission resulting from the relatively low resolution of LoLSS, which does not always allow for accurate subtraction of spurious sources.\\

\item We have measured the halo spectral index between 54 and 144 MHz for all the remaining 11 radio halos. We find that seven of them ($\sim$64\%) have an ultra-steep spectrum ($\alpha < -1.5$). This is in line with theoretical expectations from turbulent re-acceleration models, which predict an high fraction of ultra-steep spectrum halos at low frequency.\\

\item We find a tentative trend for more massive clusters to host flatter radio halos; however, given the currently poor statistics, we remind this discussion to a future work which will exploit a much larger sample.\\

\item We have discussed the limitations of our study. The current status of LoLSS and its relatively low resolution result in low statistics, which cannot conclusively determine the origin of radio halos. LoLSS DR2 will allow to significantly increase the sample, as LoTSS DR2 recently detected $\sim$80 radio halos. Combined with the increased resolution which will be provided by LOFAR international stations, it will finally elucidate the role of turbulent re-acceleration for the origin of radio halos in galaxy clusters.

\end{itemize}

\bibliographystyle{aa.bst}
\bibliography{bibliography}

\begin{appendix}
\normalsize

\section{Data calibration}
\label{app:data}

\subsection{LBA observations and calibration procedure}

The observations of our clusters were taken as part of the LoLSS survey. LoLSS is performed by observing different pointings so that the Northern Sky at DEC $>$24\degree\ is covered with a sensitivity which is close to uniform (see \citealt{deGasperin_2021} and  \citealt{deGasperin_2023} for more detail). Each pointing is independently calibrated using the automated Pipeline for LOFAR LBA (PiLL\footnote{Publicly available at \url{https://github.com/revoltek/LiLF}}). Since all the details are thoroughly discussed in \citet{deGasperin_2021} and \citet{deGasperin_2023}, here we only summarise the main steps. Phase and bandpass solutions are derived through the Default Pre-Processing Pipeline (DP3, \citealt{vanDiepen_2018}) from calibrators that are simultaneously observed with the targets (making use of the multi-beam capability of LOFAR), and transferred to the target field. Faraday rotation, second-order beam errors and direction-averaged ionospheric delays are then corrected through direction-independent calibration. To correct for direction-dependent errors, all sources are subtracted from the visibilities, and then only the brightest source is re-added (DD calibrator). Calibration solutions are derived from this source through self-calibration in DP3, then subtracting again the DD calibrator more accurately. This process is performed on every sufficiently bright source in the field of view (FoV). The FoV is then divided into facets depending on the positions of all DD sources, calibrating each facets using the solutions of the corresponding DD source. This latter process is performed through {\ttfamily{DDFacet}} \citep{Tasse_2021}.

\subsection{Extraction and imaging}

Each cluster of our sample is covered by one or more survey pointings, which have been calibrated following the standard procedure described above. However, survey images are often affected by issues such as calibration artefacts, smearing (i.e. elongation or blurring of sources) and reduction in dynamic range. These effects occur as a result of beam errors and ionospheric disturbances, that can be relevant especially in the case of wide-field images. In order to improve the image fidelity and dynamic range, we post-process LoLSS data by exploiting a strategy that has been already applied successfully to both 144 MHz observations \citep{vanWeeren_2021, Botteon_2022} and 54 MHz observations \citep[e.g.,][]{Edler_2022, Cuciti_2022, Pasini_2022}. The corresponding workflow is summarised in \citet{vanWeeren_2021} for 144 MHz data, while we already described our 54 MHz implementation in detail in \citet{Pasini_2022}. We summarise here the main steps, with the algorithm that has been suitably improved to deal with all current and future LoLSS observations. In each pointing with beam sensitivity at the target position above 30$\%$, a region is chosen around the target source. The choice is based on the flux density within this region, which typically has an extent of $15'-20'$. All sources outside this extraction region are subtracted, shifting the phase centre to the centre of the region and averaging all data in time and frequency. Pointings with higher beam sensitivity contribute more to the final, combined dataset. Self-calibration is then performed through {\ttfamily{WSClean}} \citep{Offringa_2014} and DP3. The accuracy of the flux density scale was compared to LoLSS to check for correctness. The extraction pipeline is included in the LiLF package\footnote{\url{https://github.com/revoltek/LiLF/tree/master/pipelines}}.

For each target we have produced images at nominal and high resolution, setting {\ttfamily{Briggs 0}} and {\ttfamily{Briggs -0.6}}, respectively. We also produced a number of images at low resolution. This was done by setting {\ttfamily{Briggs -0.3}}, but tapering visibilities at 30$''$, 90$''$ and to an angular scale corresponding to 50 and 100 kpc at the cluster redshift. In addition, to better highlight the diffuse emission we also subtracted compact sources using the following procedure. First, we produced an high-resolution image by setting {\ttfamily{Briggs -1}} and by cutting visibilities below a certain threshold, so that everything with a largest linear size (LLS) below 250 kpc gets removed. The angular size of the threshold depends therefore on the cluster redshift. We also tested different values, including 350 and 400 kpc, since extended AGN emission (e.g. remnant plasma from past AGN outbursts) likely constitutes the dominant contaminating factor, especially at 54 MHz, when studying diffuse emission. We found that 250 kpc works best for our purposes, since it removes the vast majority of AGN emission and leaves diffuse emission intact. The clean components of this high-resolution image, that constitute our model, are then subtracted from the overall visibilities, leaving only the diffuse emission. For these source-subtracted datasets, we separately produced multiple low-resolution images applying the same tapering discussed above. The typical \textit{rms} noise is $\sim$1.5, $\sim$2.5 and $\sim$4 mJy beam$^{-1}$ for nominal resolution, 30$''$- and 90$''$-tapered images, respectively, while tapering visibilities at 50 and 100 kpc usually leads to 1.5 and 1.7 mJy beam$^{-1}$. Images of the whole sample at both 54 and 144 MHz can be found in Appendix \ref{app:images}.

\subsection{HBA observations}

To estimate the synchrotron spectral index, we exploited the 144 MHz LoTSS observations of our clusters discussed in \citet{vanWeeren_2021}. In particular, we use the same data that was previously extracted and calibrated in \citet{Botteon_2022}\footnote{\url{https://lofar-surveys.org/planck_dr2.html}}. All targets have undergone a similar procedure to what was described above for 54 MHz, including a consistent compact source-subtraction. We refer the reader to \citet{vanWeeren_2021} and \citet{Botteon_2022} for further details.

\subsection{X-ray observations}

Since radio halos are known to be spatially and physically correlated \citep{Cassano_2010a} with thermal emission from the Intra-Cluster Medium (ICM), we have used the same X-ray observations exploited in \citet{Botteon_2022} and \citet{Zhang_2023} to check the presence of halos at 54 MHz. These are typically archival \textit{Chandra} or XMM-Newton observations, that were processed with CIAO 4.11 using CalDB v4.8.2 and SAS v16.1.0 following the standard data reduction procedure for the two instruments. We refer to \citet{vanWeeren_2021} for further detail about calibration, and to \citet{Botteon_2022} for the combination with 144 MHz observations.

\section{Initial sample}
\label{app:table}

Table \ref{tab:fulllist} provides a complete table of the initial sample of PSZ galaxy clusters analysed in this work.

\begin{table*}[h!]
\centering
\normalsize
\begin{tabular}{l l l l l l}
\hline
\hline
PSZ2 Name & Abell Name & $z$ & Mass & Included & Reason for exclusion \\
& & & [10$^{14}$ M$_\odot$] & & \\
\hline
PSZ2G080.16+57.65 & A2018 & 0.0878 & 2.51 $\pm$ 0.21 & No & Halo emission mixed with old plasma \\
PSZ2G080.70+48.31 & A2136 & 0.235 & 3.20 $\pm$ 0.41 & No & Small-scale emission (mini-halo?) \\ 
PSZ2G081.02+50.57 & / & 0.501 & 4.69 $\pm$ 0.54 & No & Complex source subtraction \\
PSZ2G084.10+58.72 & / & 0.731 & 5.40 $\pm$ 0.62 & No & High redshift \\
PSZ2G086.93+53.18 & / & 0.6752 & 5.45 $\pm$ 0.51 & Yes & / \\
PSZ2G096.83+52.49 & A1995 & 0.318 & 4.92 $\pm$ 0.37 & Yes & / \\
PSZ2G099.86+58.45 & / & 0.616 & 6.85 $\pm$ 0.49 & Yes & / \\
PSZ2G106.61+66.71 & / & 0.3314 & 4.67 $\pm$ 0.56 & No & Complex source subtraction \\
PSZ2G107.10+65.32N & A1758N & 0.2799 & 8.00 $\pm$ 0.50 & Yes & / \\
PSZ2G107.10+65.32S & A1758S & 0.2799 & 5.10 $\pm$ 0.40 & Yes & / \\
PSZ2G111.75+70.37 & A1697 & 0.183 & 4.34 $\pm$ 0.33 & Yes & / \\
PSZ2G112.48+56.99 & A1767 & 0.070 & 2.99 $\pm$ 0.15 & No & Complex source subtraction \\
PSZ2G114.31+64.89 & A1703 & 0.2836 & 6.76 $\pm$ 0.37 & Yes & / \\
PSZ2G118.34+68.79 & / & 0.2549 & 3.77 $\pm$ 0.49 & No & Complex source subtraction \\
PSZ2G133.60+69.04 & A1550 & 0.254 & 5.88 $\pm$ 0.40 & Yes & / \\
PSZ2G135.17+65.43 & / & 0.5436 & 6.01 $\pm$ 0.60 & Yes & / \\
PSZ2G139.18+56.37 & A1351 & 0.322 & 6.87 $\pm$ 0.38 & Yes & / \\
PSZ2G143.26+65.24 & A1430 & 0.3634 & 7.65 $\pm$ 0.43 & Yes & / \\
PSZ2G150.56+58.32 & / & 0.466 & 7.55 $\pm$ 0.51 & No & Complex source subtraction \\
PSZ2G156.26+59.64 & / & 0.6175 & 6.77 $\pm$ 0.60 & No & Complex source subtraction \\
\hline
\hline
\end{tabular}
\caption{\footnotesize The table lists our initial sample of PSZ2 galaxy clusters hosting either confirmed or candidate radio halos, following the classification of \citep{Botteon_2022}, together with their Abell name, redshift and mass. The last two columns report whether the cluster was included in the final sample and, if not, the main reason for the exclusion.} 
\label{tab:fulllist} 
\end{table*} 

\section{Flux density and spectral index measurement}
\label{app:calculations}

In order to calculate the spectral indices of these 11 halos, we measured their flux density at 54 and 144 MHz. This was done for each system at the highest resolution at which the combination of image quality, source subtraction and halo visibility was best. To measure the halo flux density, recent studies \citep[e.g.,][]{Osinga_2021, Botteon_2022, Bruno_2023} have exploited the Halo-FDCA algorithm \citep{Boxelaar_2021}, which fits the surface brightness radial profile of radio halos with an exponential function. However, the low S/N ratio of most radio halos at 54 MHz does not allow a reliable fit with this procedure. Furthermore, the algorithm is currently not able to simultaneously fit the surface brightness at the two frequencies (see \citealt{Botteon_2022} for a discussion of the algorithm limitations). The latter restriction impacts our analysis since performing independent fits at the two frequencies would likely lead us to sample (slightly) different areas of diffuse emission, which would translate into an erroneous spectral index value. 

Hence, the calculation of the halo flux density at each frequency is done from circular regions centered on the peak of the halo as detected at 144 MHz in \citet{Botteon_2022}\footnote{Spurious sources which are not part of the halo emission (e.g. leftovers of the source subtraction, old and diffuse AGN plasma) were excluded from the flux density calculation when clearly distinguishable.}, with a radius corresponding to one e-folding radius ($r_\mathrm{e}$) as estimated at 144 MHz. This is motivated by the fact that $r_\mathrm{e}$, which describes the size of the radio halo through an exponential profile $I(r)\propto e^{-r/r_\mathrm{e}}$, is defined as the radius at which the halo surface brightness goes down by a factor 1/\textit{e} (and thus is roughly 1/3 of the emission peak). Therefore, it yields a constant fraction $S_{r_\mathrm{e}}$ $\sim$ 30\% of the total flux density \citep[e.g.,][]{Bruno_2023, Cassano_2023}. Typically, 3$r_\mathrm{e}$ is used as a reference, which recovers $\sim$80\% of the total flux density. However, we decided to use 1$r_\mathrm{e}$ in order to exclude as much as possible any kind of contamination by spurious sources. In some of our systems, using 3$r_\mathrm{e}$ would have implied to make a wide use of masks, which we tried to avoid in order to provide a consistent, easily-reproducible measurement. Even though we might miss halo flux in the outskirts, it should still provide accurate estimates of the integrated spectral index. Currently, there are only a few known cases of radio halos showing gradients in the synchrotron spectrum steepness moving from the halo centre towards the periphery. It is worth noting that one of these is the halo hosted in Coma, which is the closest known to-date \citep{Brunetti_2001, Bonafede_2022}. Nevertheless, we expect the integrated spectral index to be barely affected by any gradient, as it is weighted by the flux density.

Uncertainties on the flux density, $S$, are then derived using:

\begin{equation}
 \sigma_S^2 = N_{\rm{beams}}\sigma_{\rm{rms}}^2 + \sigma_{\rm{sub}}^2 +  \left(f\times S\right)^{2} \mbox{ ,}
\end{equation}
where $f=0.1$ for 144 MHz data and $f=0.06$ for 54 MHz data is the absolute flux-scale uncertainty \citep{Shimwell_2019, Shimwell_2022, deGasperin_2023}, $N_{\rm{beams}}$ the number of beams covering the halo (within 1 $r_\mathrm{e}$), $\sigma_{\rm{rms}}$ the image noise, and $\sigma_{\rm{sub}}$ the uncertainty due to compact source subtraction. The latter is given by:
\begin{equation}
   \sigma_{\rm{sub}}^2 = \sum_{i}
   N_{\rm{beams,i}}\sigma_{\rm{rms}}^2  \mbox{ ,}
\end{equation}
where the sum is taken over all the $i$ sources that were subtracted within the region in which the flux density is estimated. 

Residual images were carefully inspected to check for the presence of non-deconvolved emission, which could lead to overestimating the flux density. For all our systems, the flux density within the same region (i.e. a circle with radius = 1$r_\mathrm{e}$) as measured from residual images was found to be consistent with 0, as expected.

In addition, as an independent check, we have also calculated the spectral index by measuring the flux density of each halo at the two frequencies within the 3$\sigma$ contours derived at 144 MHz. With this method we might mistakenly include spurious emission that was not completely subtracted by our procedure, but for the majority of the sample we should still be able to get a good spectral index estimate. With this alternative measurement, our aim is to confirm that using 1$r_\mathrm{e}$ did not yield any bias due, for example, to the relatively small dimensions of the region.  Finally, we checked the literature for the few systems for which a spectral index (within the same or a different frequency interval) had already been measured \cite{Botteon_2020, diGennaro_2021, Hoeft_2021, Pasini_2022}. We find consistent values for the spectral index with all three methods, the only exception being PSZ2G099.86+58.45. In this system, ref. \cite{diGennaro_2021} reports an overall integrated spectral index of $\alpha_{144}^{650} = 1.00 \pm 0.13$ (measured on the whole halo extent). With our data, we find instead $\alpha = -1.67 \pm 0.26$ between 54 and 144 MHz (within 1$r_\mathrm{e}$). The discrepancy is most likely due to the combination of relatively high redshift (\textit{z} =0.616), complex morphology of the source and large number of radio AGN located within the halo (see also Fig. A.4 of ref. \cite{diGennaro_2021}), which all make the subtraction of compact sources significantly harder, especially at 54 MHz where the S/N is lower and the image beam larger. It is also worth noting that ref. \cite{diGennaro_2021} reports regions with a significantly steeper spectral index ($\alpha \sim -1.6$) in the central part of the halo. Albeit deeper observations would be beneficial to shed more light on this issue, we choose to still include the system in our sample, in order not to introduce a selection effect.

Finally, the spectral index and associated error were estimated by using: 

\begin{equation}
\label{eq:spindex}
\alpha_{\nu1}^{\nu2} = \dfrac{\ln S_1 - \ln S_2}{\ln \nu_2 - \ln \nu_1} \pm \dfrac{1}{\ln \nu_2 - \ln \nu_1} \sqrt{\bigg(\frac{\sigma_1}{S_1}\bigg)^2 + \bigg(\frac{\sigma_2}{S_2}\bigg)^2} ,
\end{equation}
where $S_1$ and $S_2$ are the flux densities at frequencies $\nu_1$ and $\nu_2$, respectively, while $\sigma$ is the corresponding error.

\section{Selection bias}
\label{app:selection}

\begin{figure}[b!]
\hspace{0.4cm}
{\includegraphics[scale=0.34]{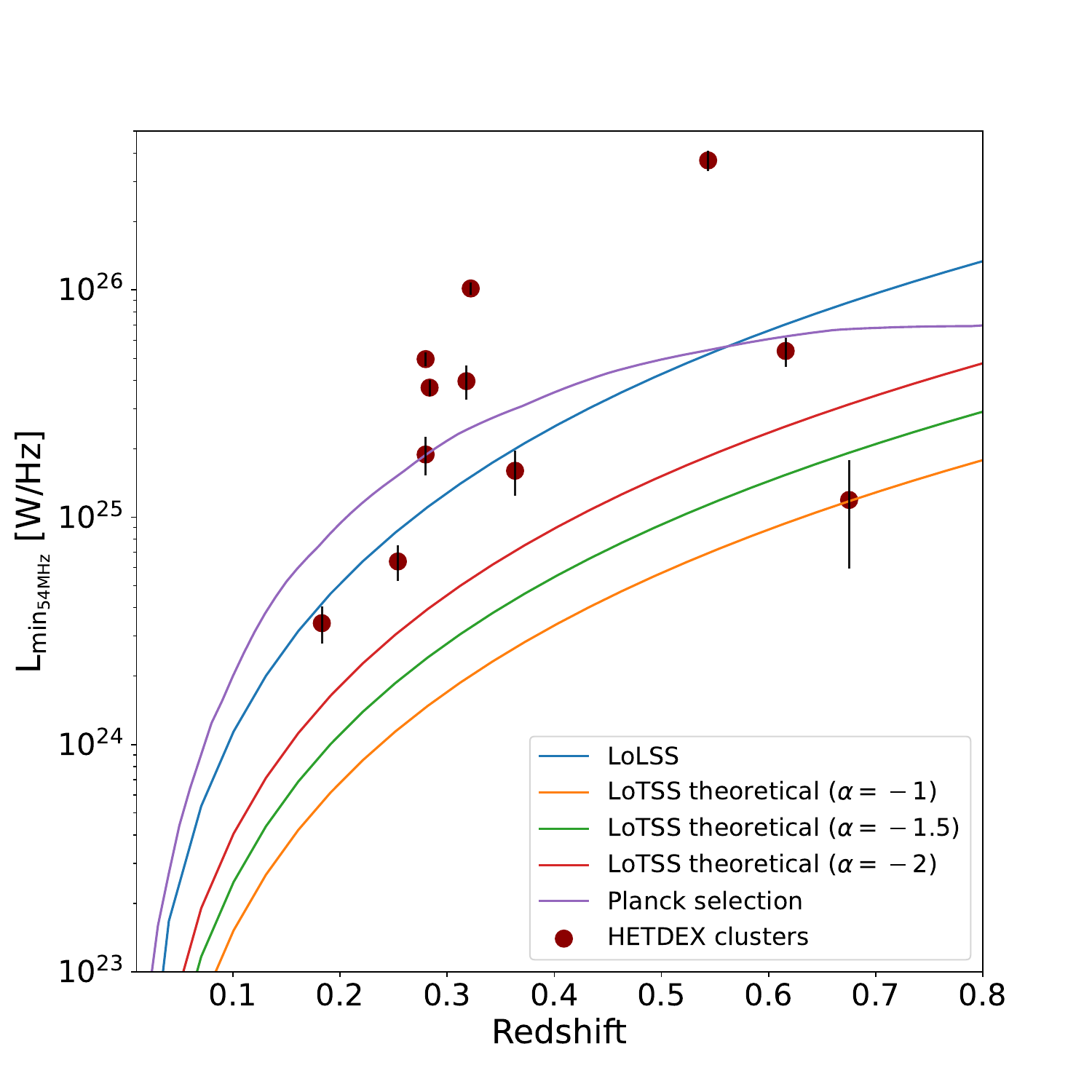}}
\caption{\footnotesize Radio power at 54 MHz of radio halos vs. redshift (red circles). The purple curve represents the 50\% PSZ2 completeness line converted to 54 MHz radio power following the power-mass correlation of \citep{Cassano_2023} and assuming $\alpha = -1.5$. The blue curve is the minimum luminosity detectable at 54 MHz by LOFAR following Eq. 2 of \citep{Cassano_2023}, estimated assuming $r_\mathrm{e} =$ 150 kpc and a resolution of 100 kpc. The red, green and yellow curves report instead the minimum luminosity detectable at 144 MHz by LOFAR under the same assumptions, converted to 54 MHz assuming $\alpha = -2, -1.5$ and $-1$, respectively.}
\label{fig:check}
\end{figure}

In this section, we discuss about the possibility that the detection of a significant population of USSRH in Fig.~\ref{fig:hist} might be driven by selection effects that originate from the different sensitivity of LoTSS and LoLSS at their respective frequency. First of all, a straightforward calculation (see also \citealt{deGasperin_2023}) shows that, if we take into account the reported sensitivity of the two surveys ($\sim$0.1 and $\sim$1 mJy beam$^{-1}$ at 144 and 54 MHz, respectively), and we rescale them to the same beam, the spectral index between them is $\sim$-0.5. This means that sources that are on the edge of LoTSS sensitivity might be not detected in LoLSS if their spectrum is flatter than this value, since their emission at 54 MHz would fall below its sensitivity limit. On the other hand, we should detect all sources with a spectrum steeper than -0.5, including all radio halos.

We have also compared the radio power of our halos as a function of the redshift with the sensitivity curves of LoTSS and LoLSS. First we have used Eq. 2 from \citep{Cassano_2023} to derive the minimum luminosity detectable at 54 MHz by LoLSS by assuming $r_\mathrm{e}$ = 150 kpc\footnote{rescaled to angular size at a given redshift, and correspondent to $\theta_{e(z)}$ in Eq. 2.} at a resolution of 100 kpc (converted in arcsec at the cluster redshift), which is the typical resolution of the images that we have used to detect radio halos. We have assumed a conservative \textit{rms} noise of 3 mJy/beam, which is the highest noise we find among the images of our halos at that resolution. We have done the same for LoTSS, converting the 144 MHz luminosity to 54 MHz by assuming different spectral indices: -1, -1.5 and -2. Finally, we have sampled the PSZ2 selection effect, similarly to what was done in \citep{Cassano_2023}. We estimated the 150 MHz radio power corresponding to the 50\% PSZ2 completeness curve ($M$, $z$) in Fig. 1 of \citep{Cassano_2023}, using the radio power-cluster mass correlation found for LoTSS-DR2 radio halos (see Eq. 1 of \citep{Cassano_2023}). We then converted the 150 MHz radio power to 54 MHz by assuming a spectral index $\alpha = -1.5$. The result is shown in Fig. \ref{fig:check}.  

The plot clearly shows that, within the redshift range of our clusters (0.18 $< z <$ 0.7), the detection of radio halos is not dominated by the sensitivity of either LoTSS or LoLSS, as the vast majority of our systems lie above or is consistent with their minimum luminosity curves. This is also supported by the fact that all our radio halos detected in LoTSS are also observed in LoLSS. In fact, the distribution of our clusters in the radio power-redshift plane of Fig.~\ref{fig:check} is determined by the PSZ2 completeness curve, which dominates over the sensitivity of the two surveys. There is one system (PSZ2G086.93+53.18) which lies well below the sensitivity curves. However, the corresponding images are characterized by a lower \textit{rms} noise than the the sensitivity curve ($\sim$3 mJy/beam), and it is the only halo that we detected by tapering visibilities to 90$''$. We can thus conclude that our results are not substantially driven by selection effects that could prevent the detection of radio halos with different spectral properties, such as the LOFAR sensitivity at the two frequencies.

\onecolumn
\section{Images}
\label{app:images}

In this section we show high- and source-subtracted low-resolution images at 54 and 144 MHz of our sample of radio halos. For each target, the low-resolution images are at the resolution from which the halo flux density was estimated. Images of the whole initial sample of clusters, as well as X-ray maps, can be found at \url{https://lofar-surveys.org/planck_dr2.html}.

\begin{figure}[h!]
\centering`

{\includegraphics[scale=0.18]{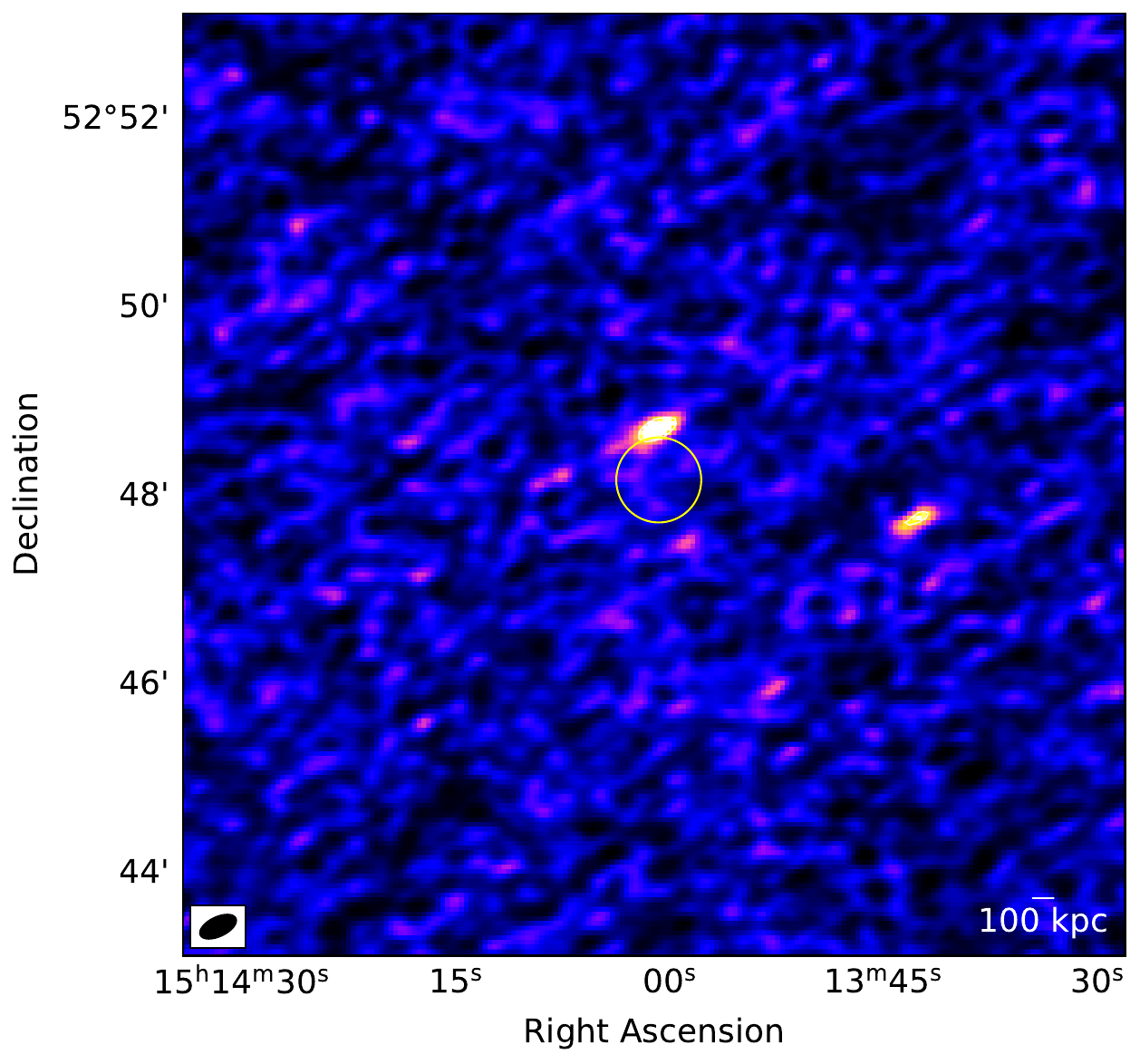}}
{\includegraphics[scale=0.18]{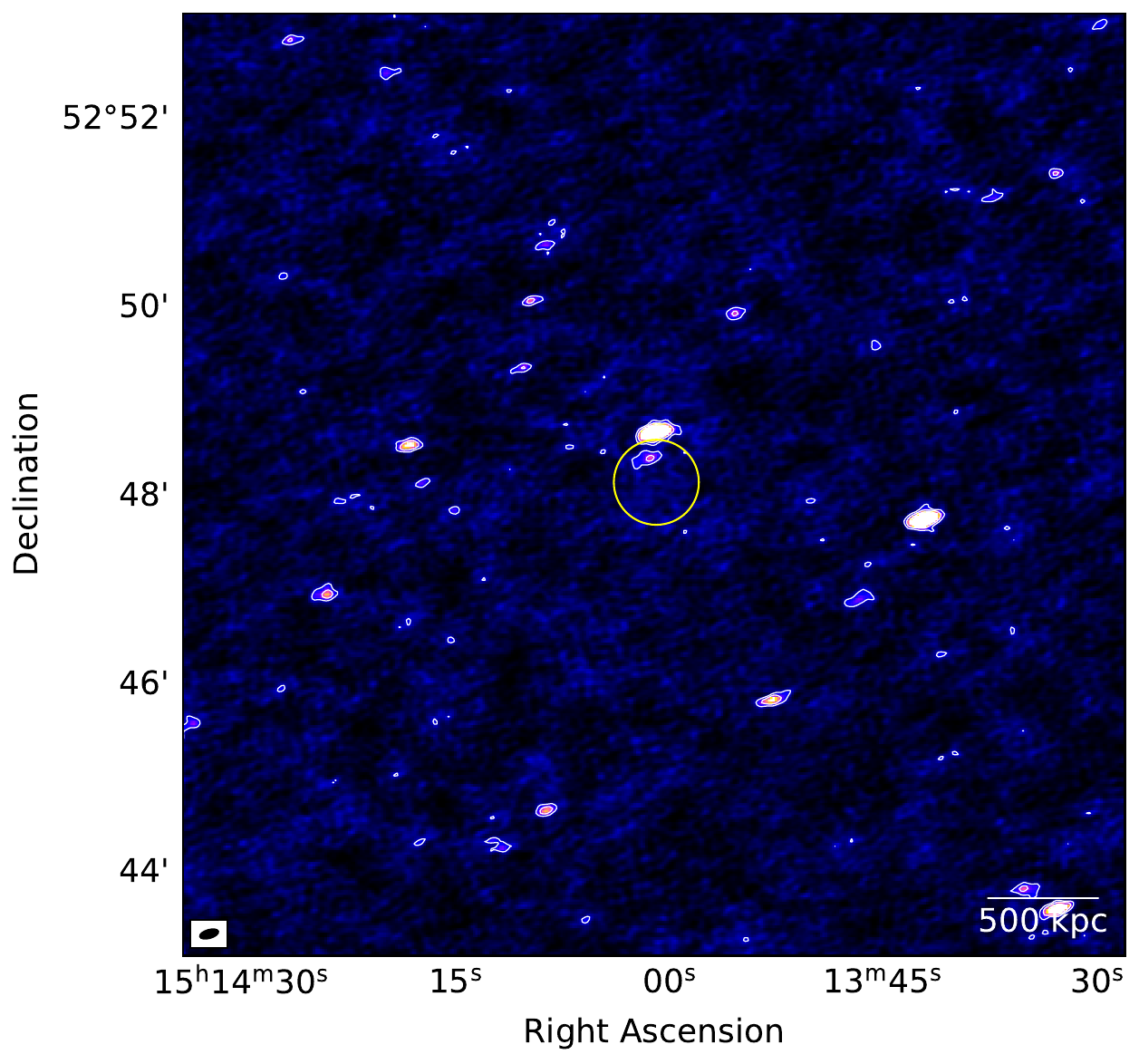}}
{\includegraphics[scale=0.18]{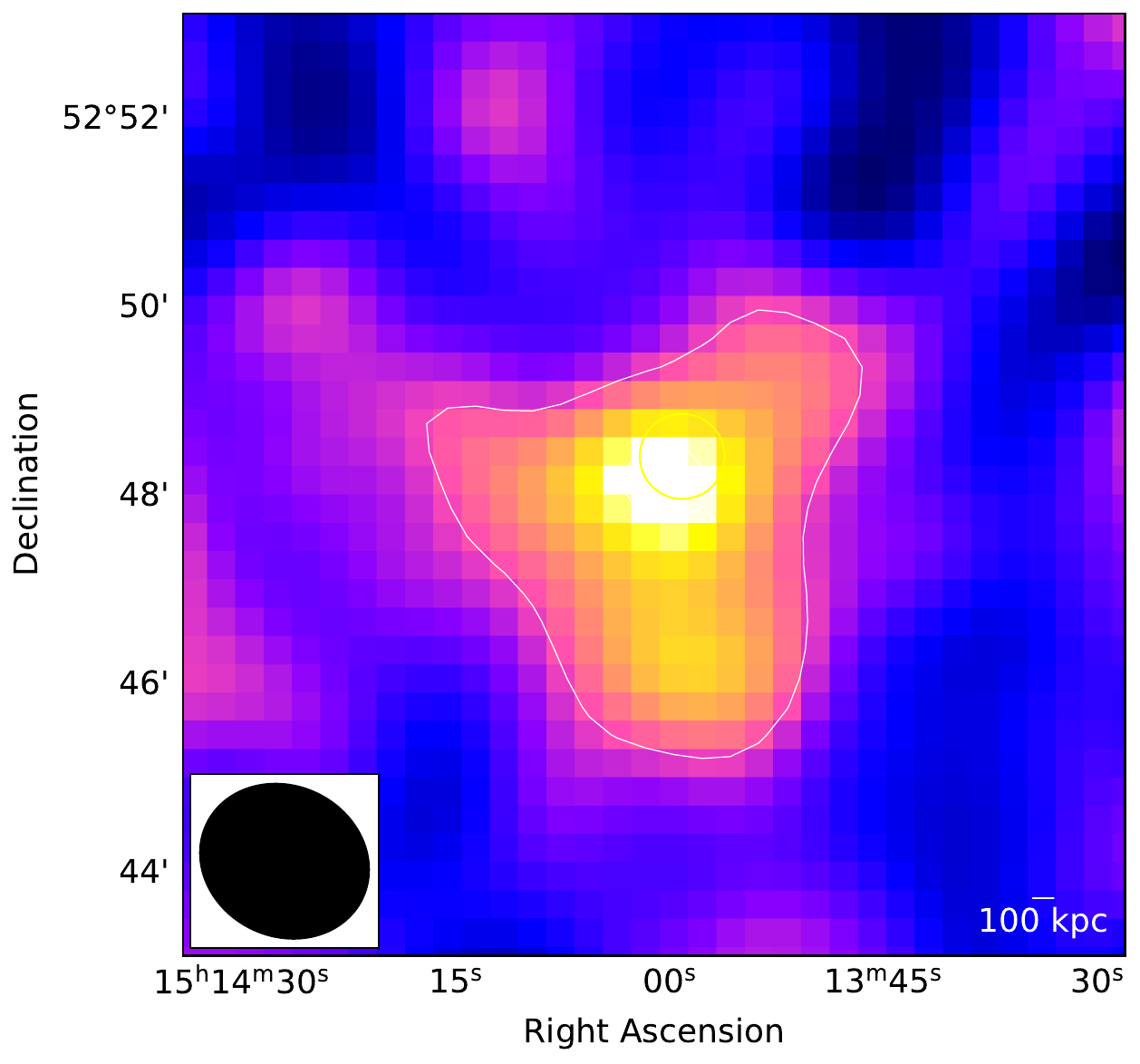}}
{\includegraphics[scale=0.18]{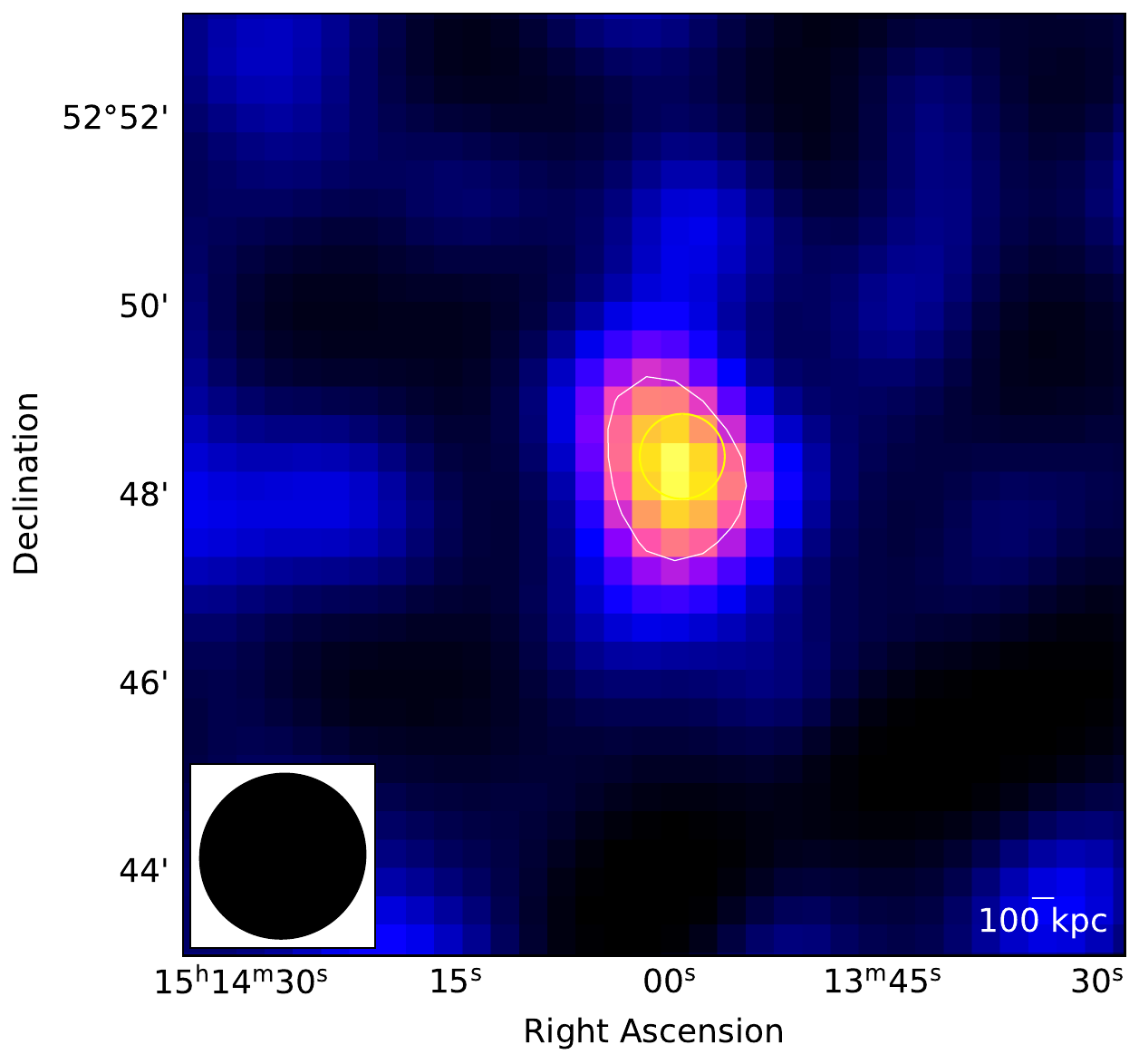}}
\caption{\footnotesize \textit{From left to right}: 54 MHz briggs, 144 MHz briggs, 54 MHz low-resolution and 144 MHz low-resolution images of PSZ2G086.93+53.18. Low-resolution images were produced by tapering visibilities at 90$''$. Their \textit{rms} noise is $\sim$5 and $\sim$1 mJy beam$^{-1}$ at 54 and 144 MHz, respectively. The yellow circle denotes 1 $r_\mathrm{e}$.}
\label{fig:PSZ086}
\end{figure}

\begin{figure}[h!]
\centering
{\includegraphics[scale=0.18]{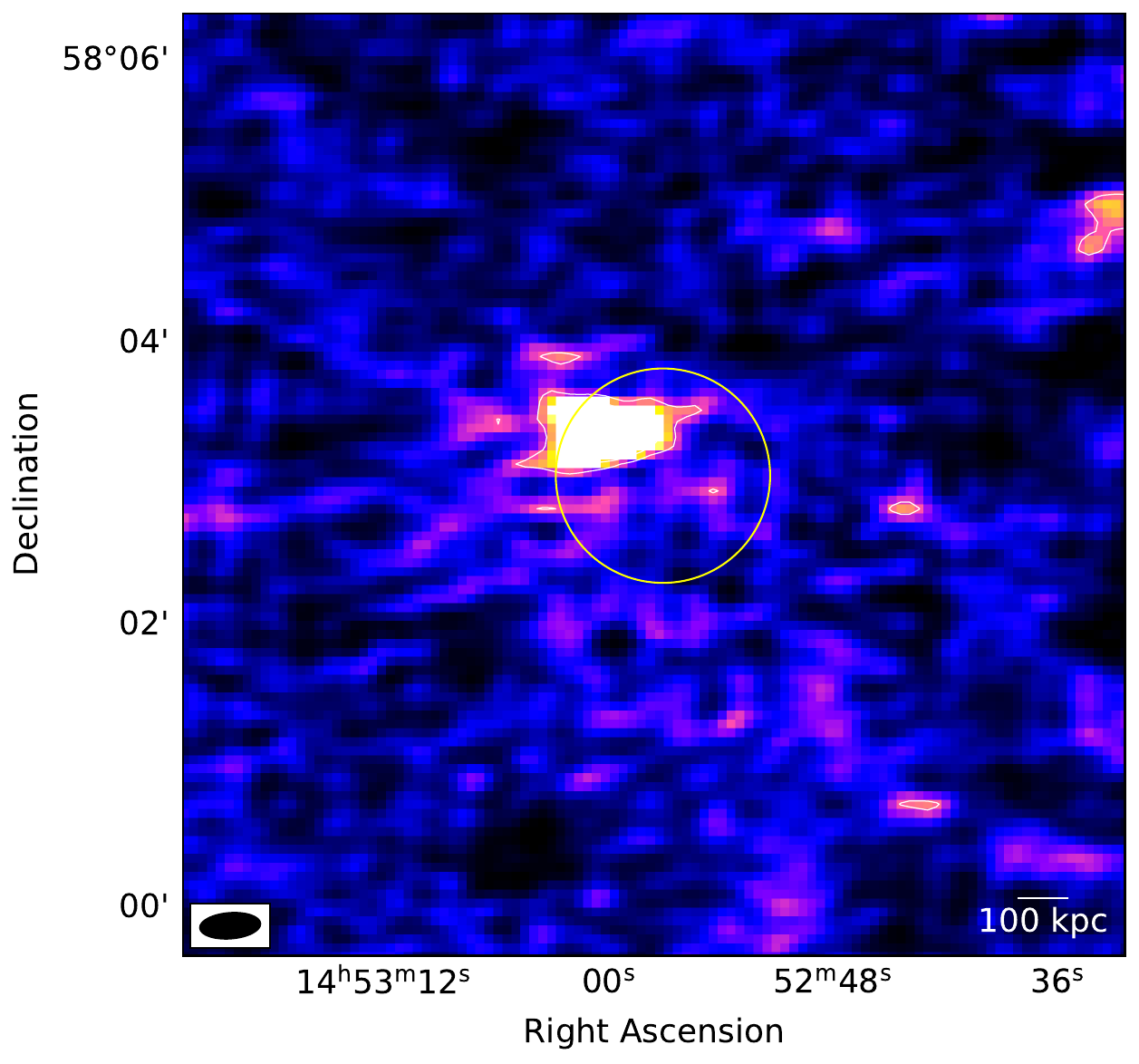}}
{\includegraphics[scale=0.18]{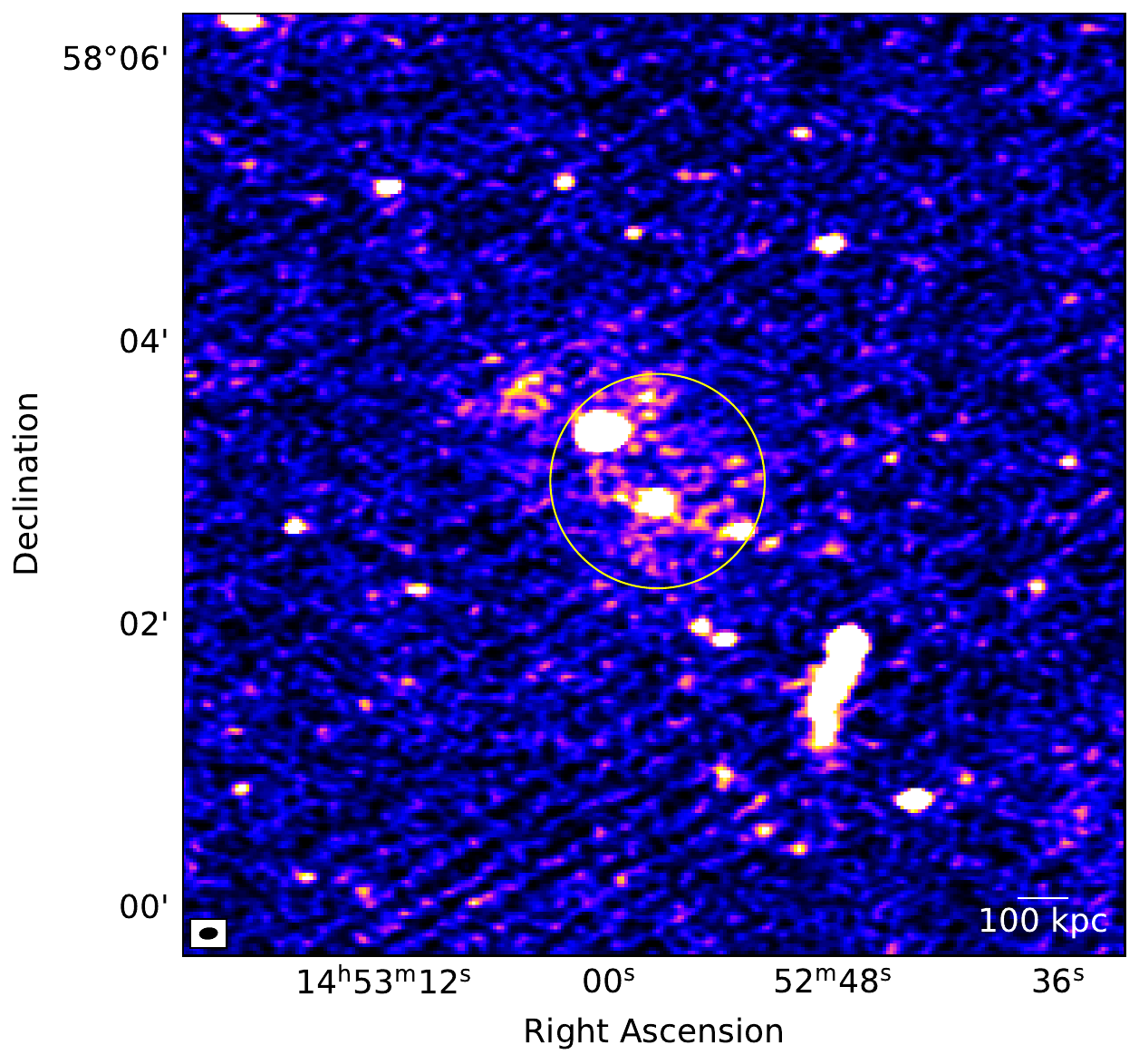}}
{\includegraphics[scale=0.18]{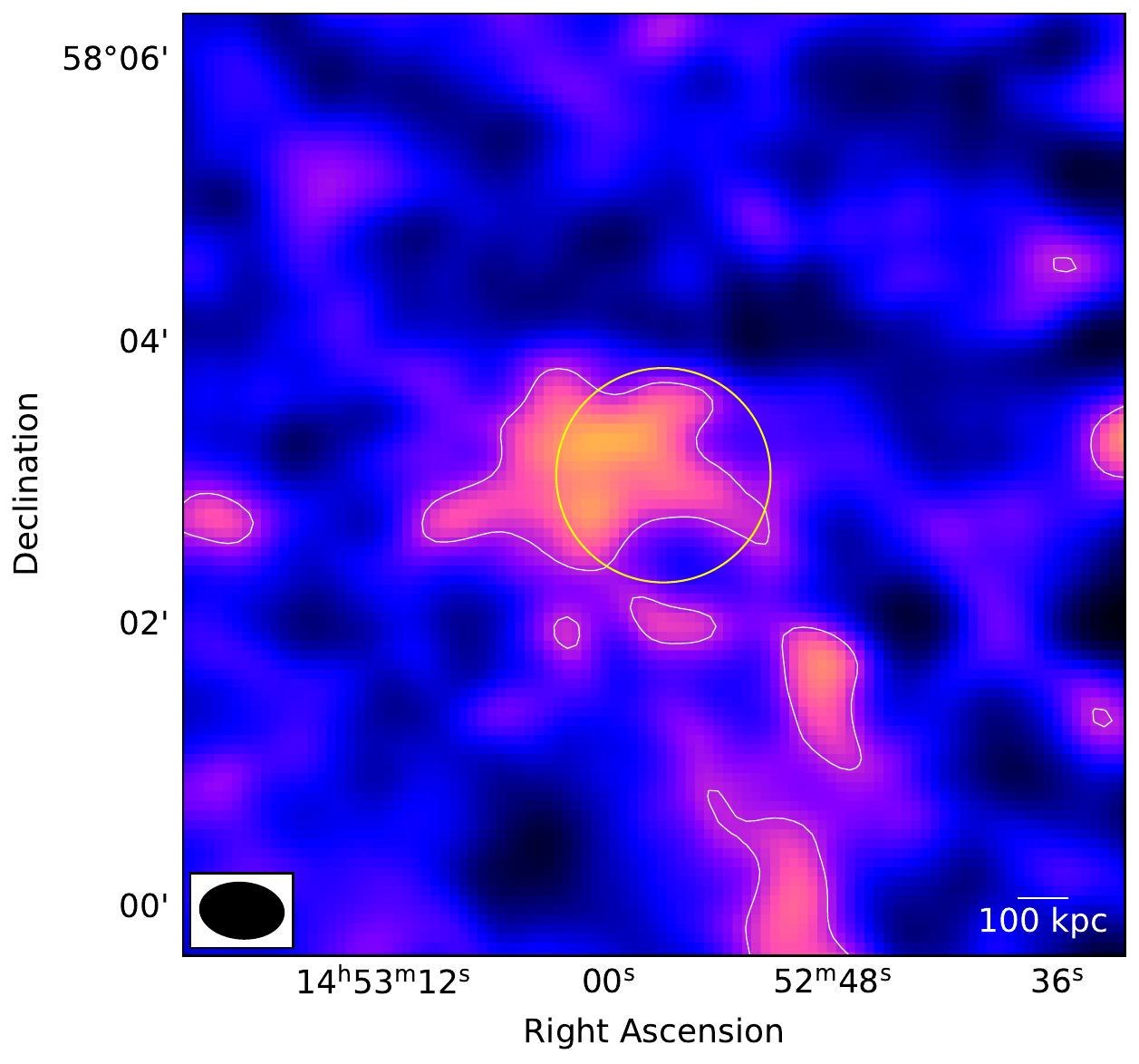}}
{\includegraphics[scale=0.18]{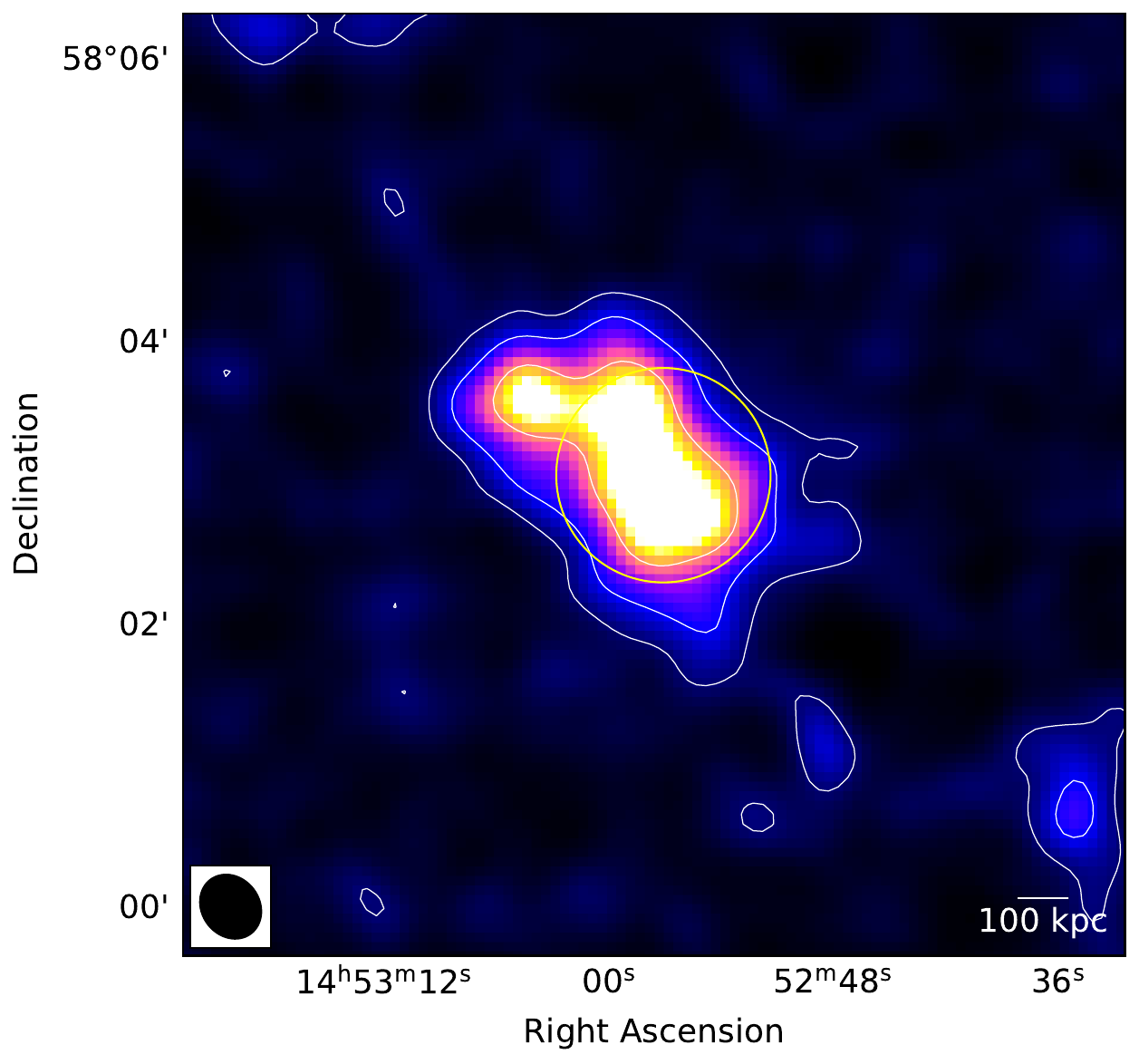}}
\caption{\footnotesize \textit{From left to right}: 54 MHz briggs, 144 MHz briggs, 54 MHz low-resolution and 144 MHz low-resolution images of PSZ2G096.83+52.49. Low-resolution images were produced by tapering visibilities at an angular scale corresponding to 100 kpc at the cluster redshift. Their \textit{rms} noise is $\sim$4 and $\sim$0.15 mJy beam$^{-1}$ at 54 and 144 MHz, respectively. The yellow circle denotes 1 $r_\mathrm{e}$.}
\label{fig:PSZ096}
\end{figure}

\begin{figure}[h!]
\centering
{\includegraphics[scale=0.18]{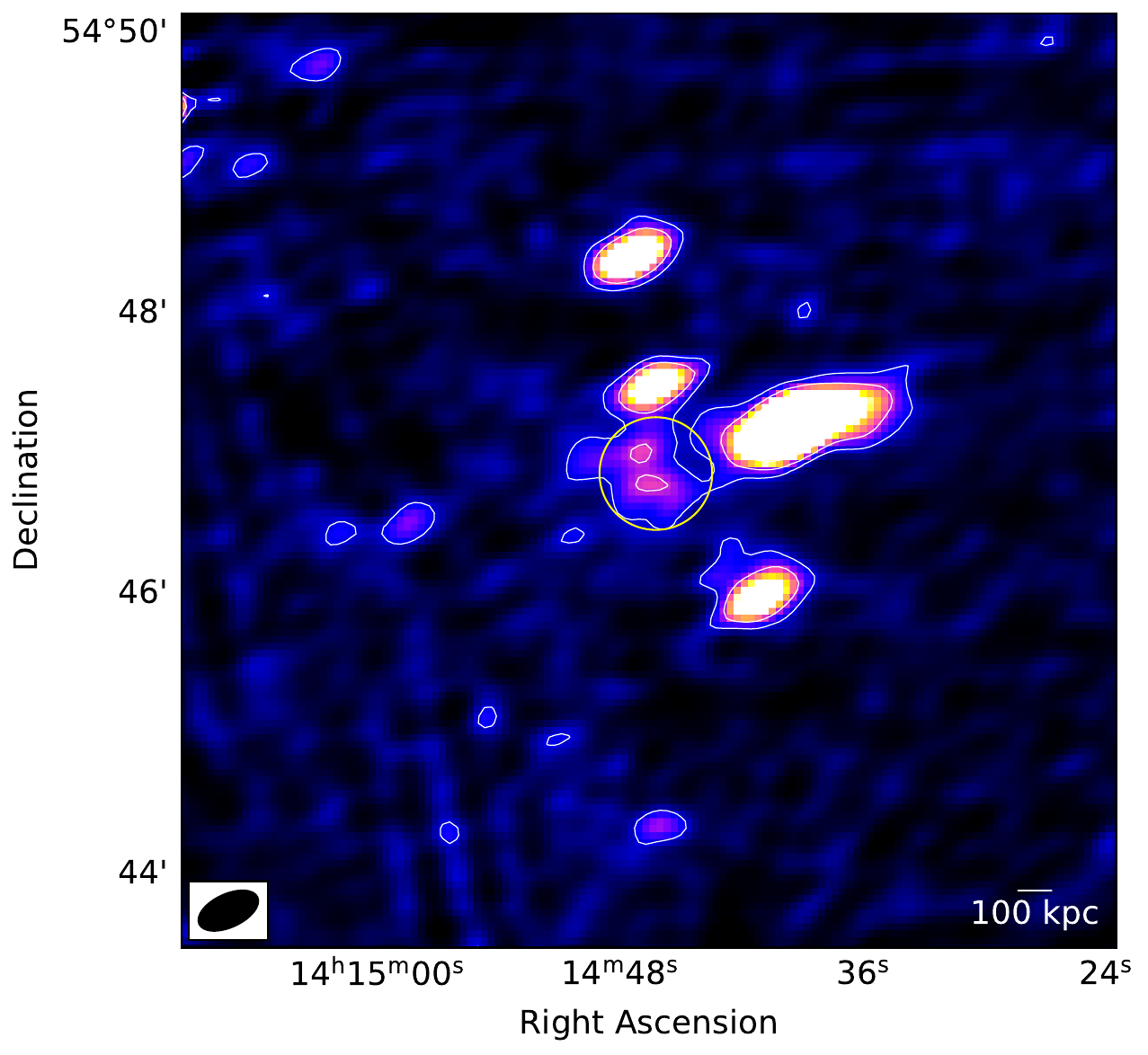}}
{\includegraphics[scale=0.18]{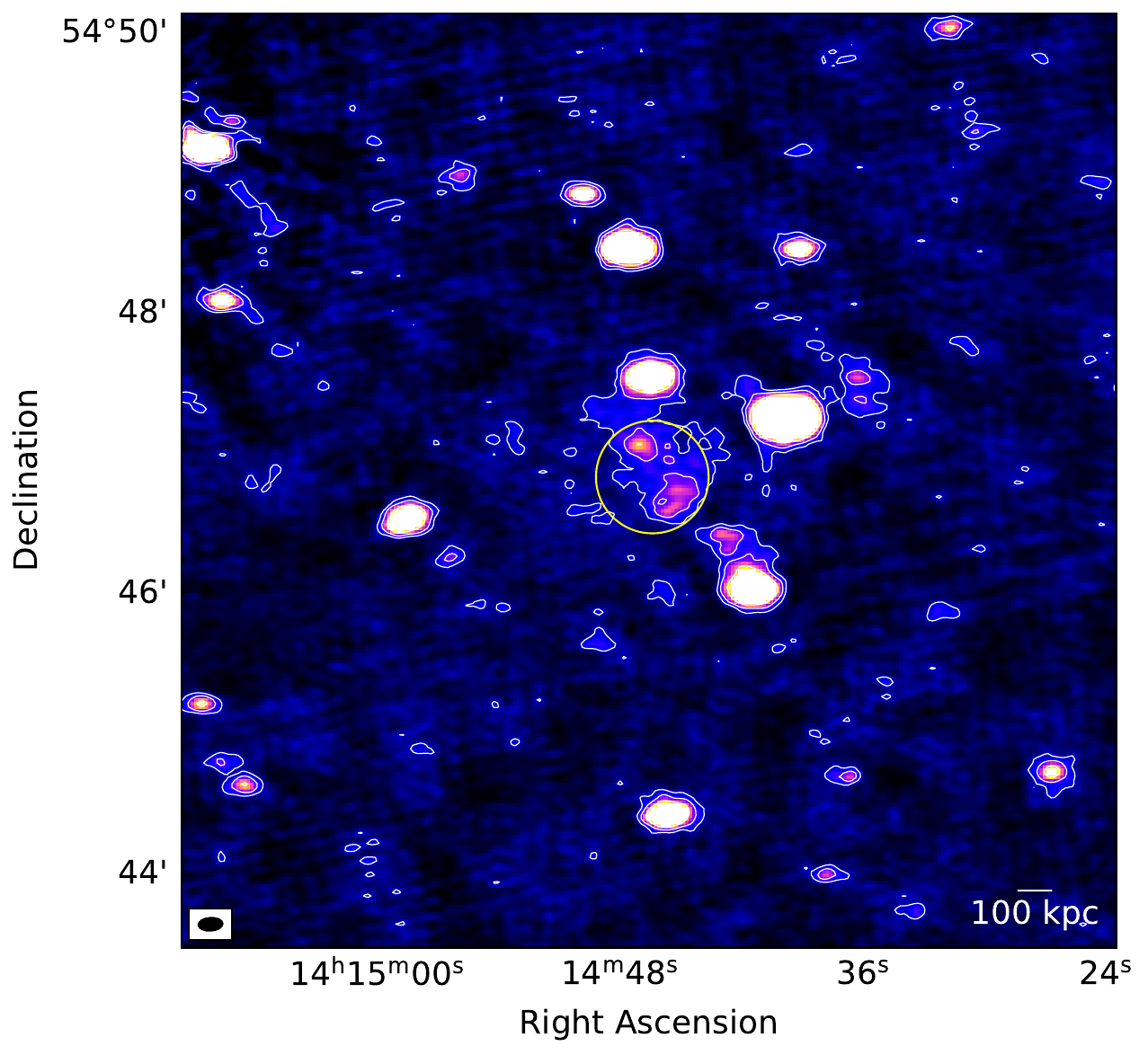}}
{\includegraphics[scale=0.18]{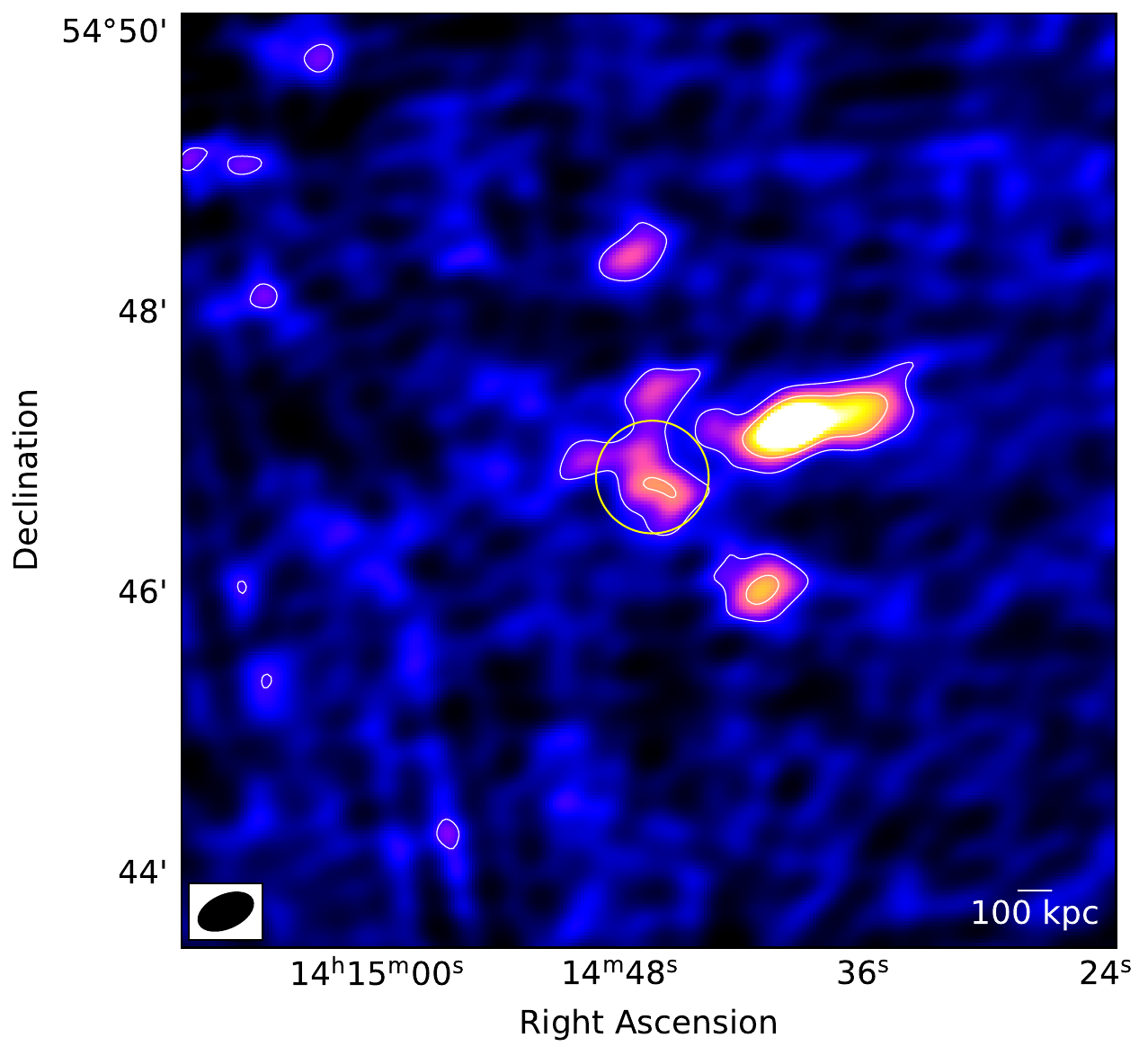}}
{\includegraphics[scale=0.18]{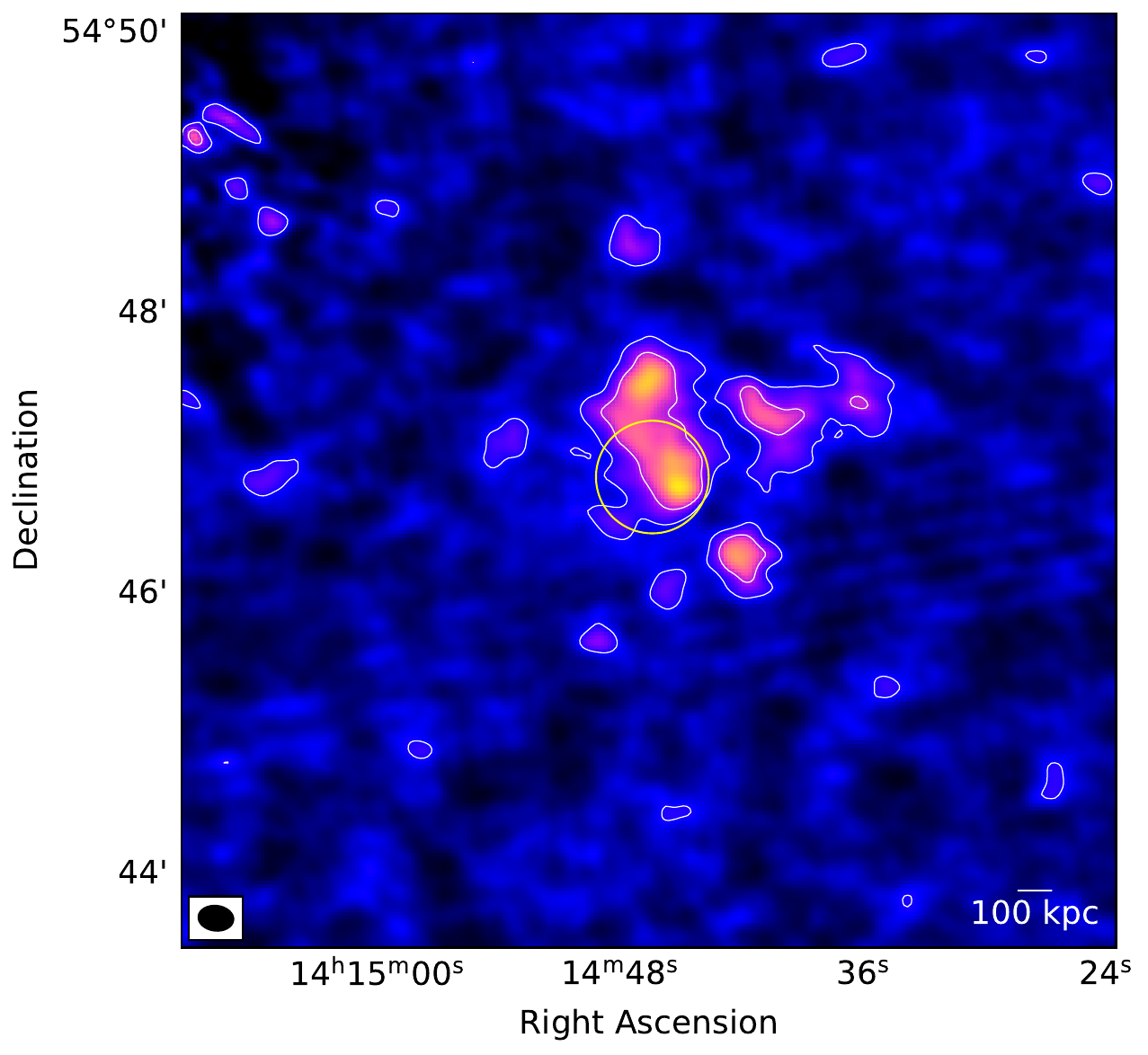}}
\caption{\footnotesize \textit{From left to right}: 54 MHz briggs, 144 MHz briggs, 54 MHz low-resolution and 144 MHz low-resolution images of PSZ2G099.86+58.45. Low-resolution images were produced by tapering visibilities at an angular scale corresponding to 50 kpc at the cluster redshift. Their \textit{rms} noise is $\sim$1.7 and $\sim$0.1 mJy beam$^{-1}$ at 54 and 144 MHz, respectively. The yellow circle denotes 1 $r_\mathrm{e}$.}
\label{fig:PSZ099}
\end{figure}

\begin{figure}[h!]
\centering
{\includegraphics[scale=0.18]{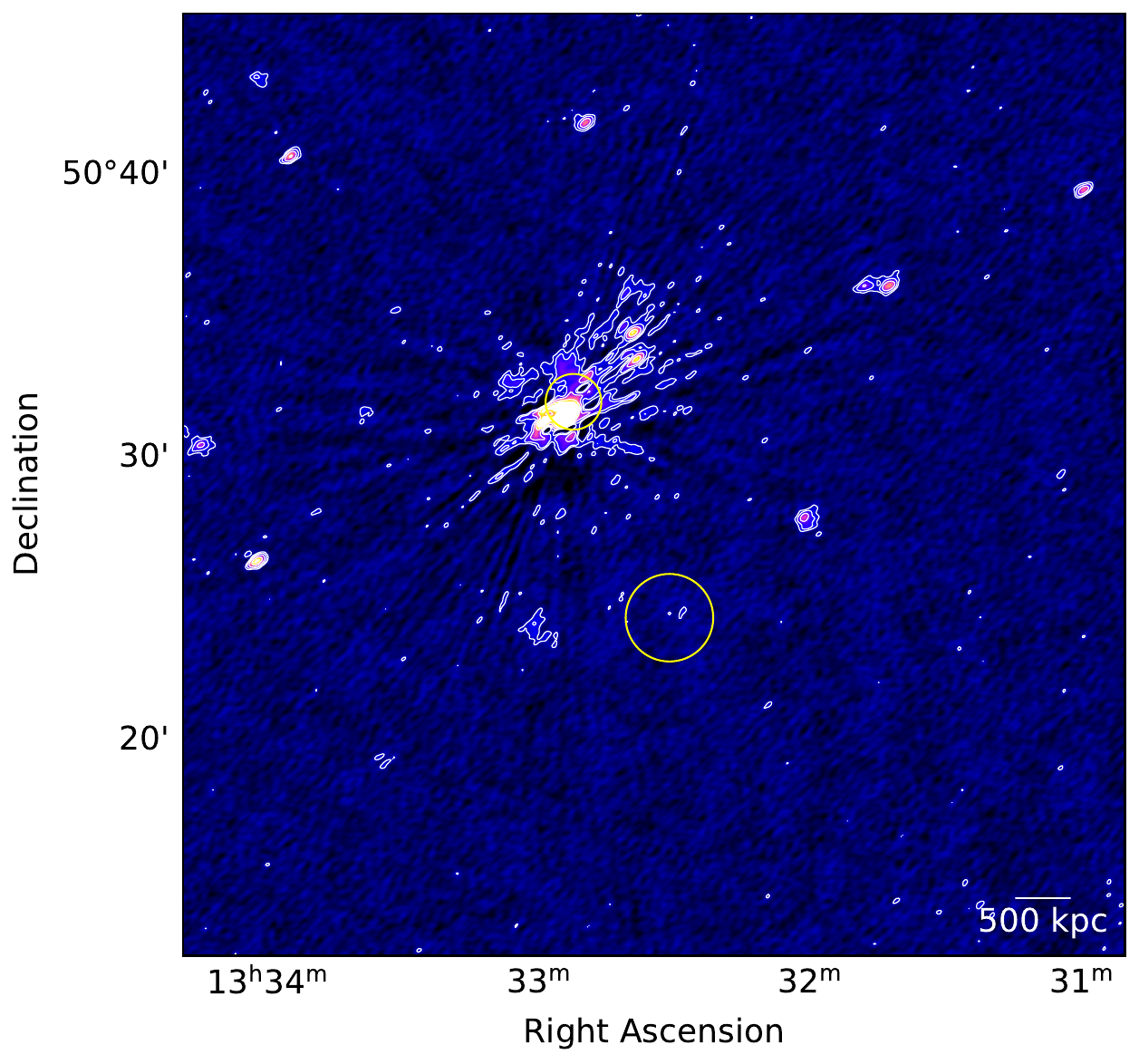}}
{\includegraphics[scale=0.18]{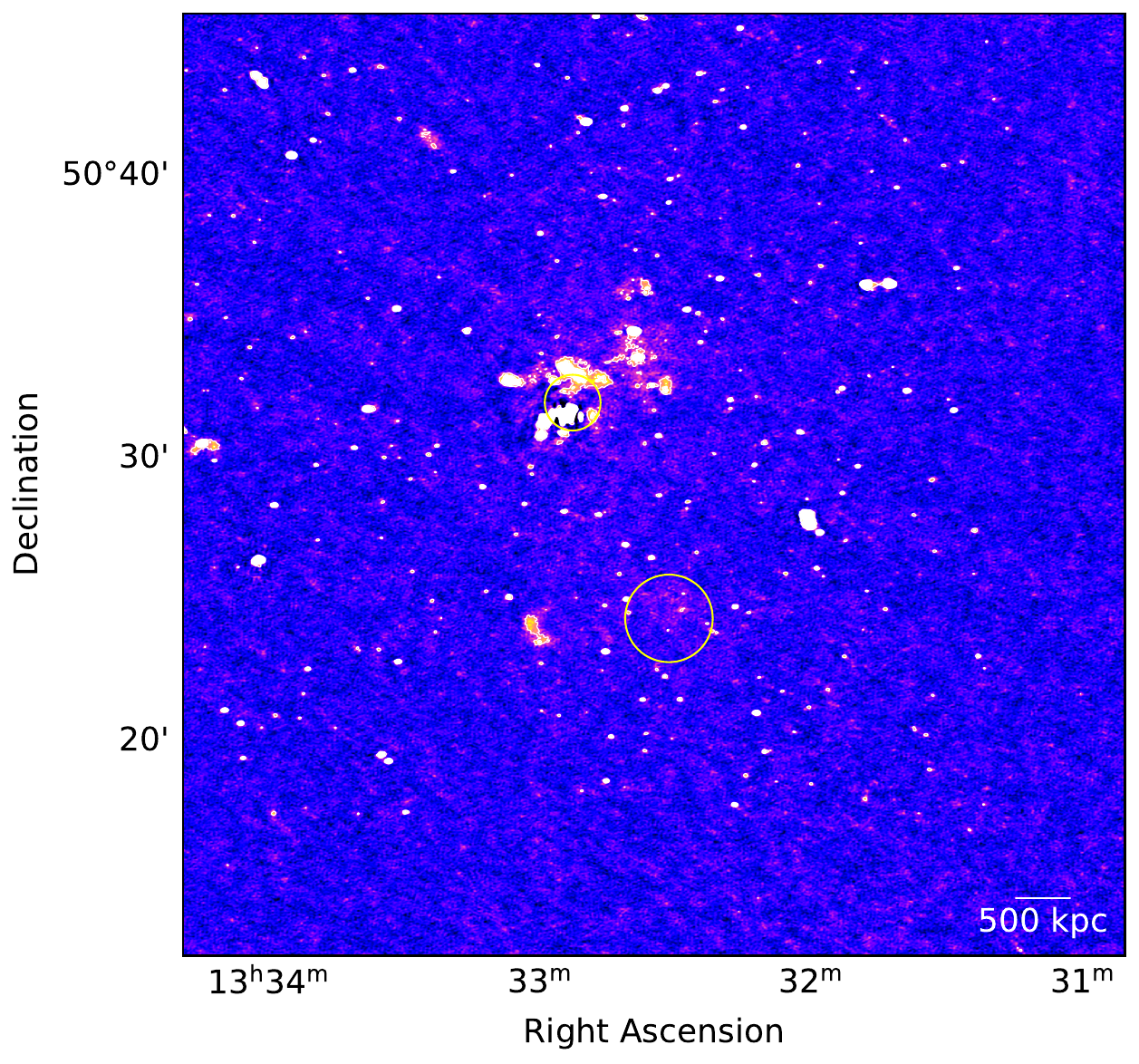}}
{\includegraphics[scale=0.18]{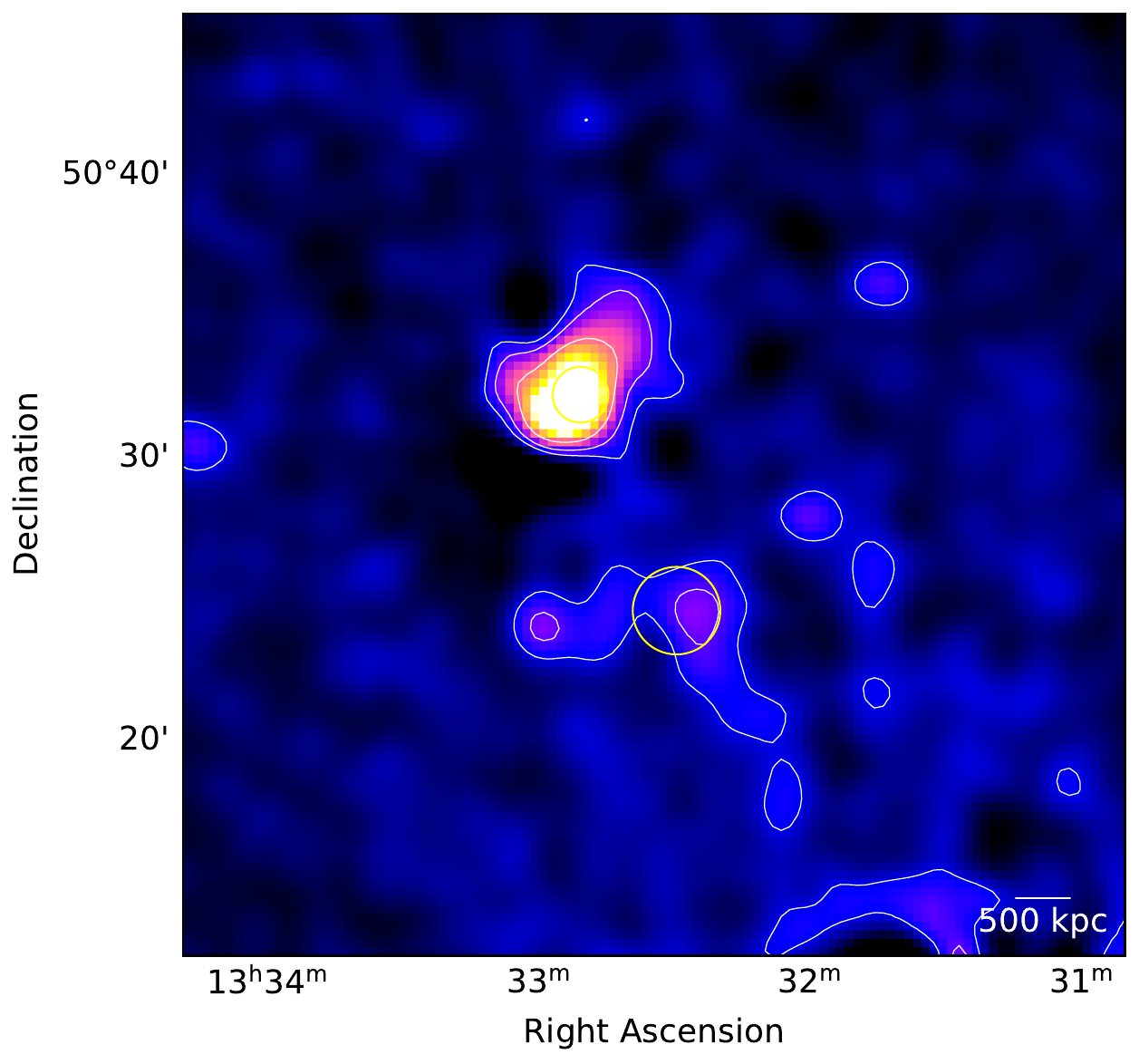}}
{\includegraphics[scale=0.18]{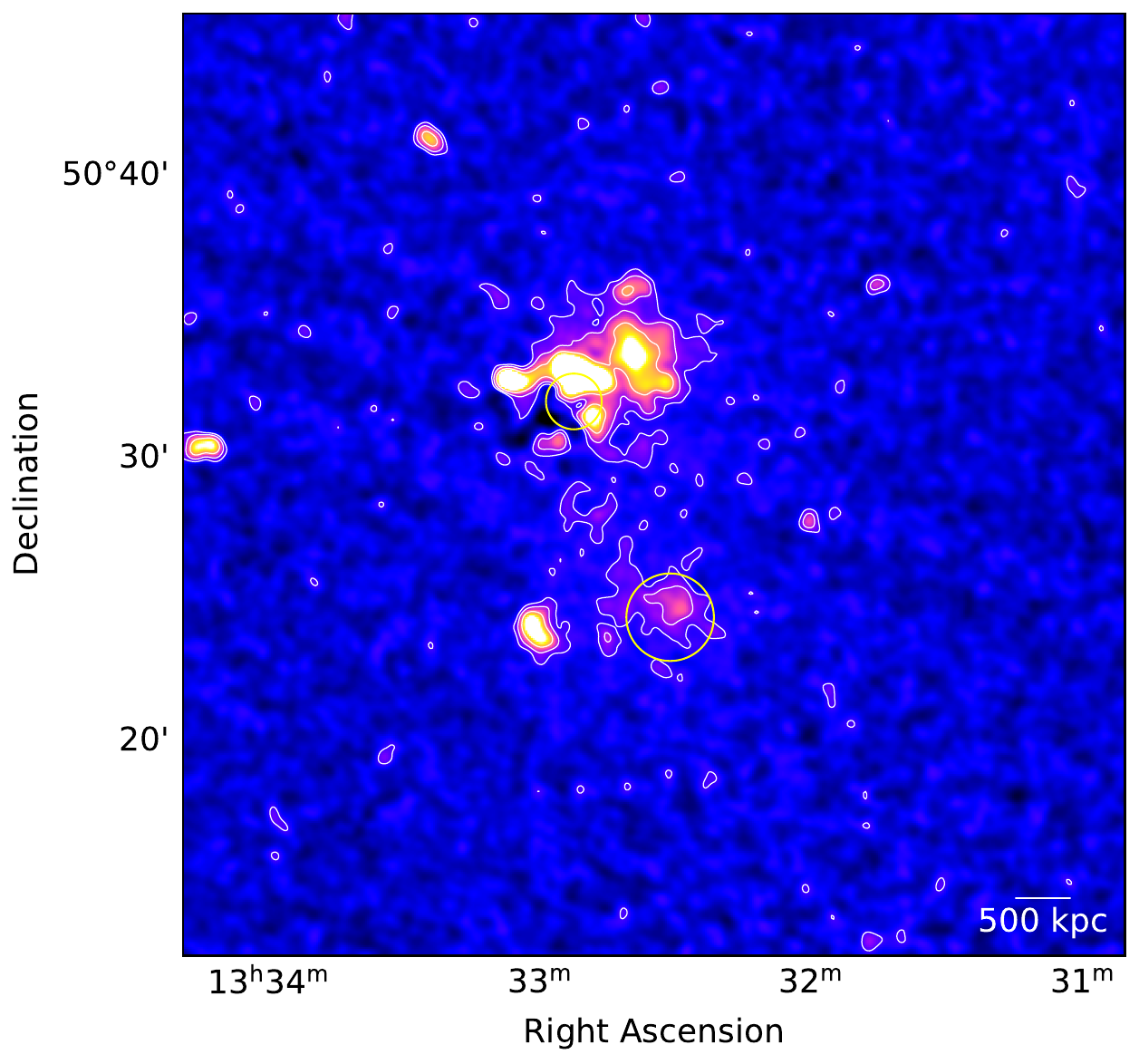}}
\caption{\footnotesize \textit{From left to right}: 54 MHz briggs, 144 MHz briggs, 54 MHz low-resolution and 144 MHz low-resolution images of PSZ2G107.10+65.32. Low-resolution images were produced by tapering visibilities at an angular scale corresponding to 100 kpc at the cluster redshift. Their \textit{rms} noise is $\sim$2.5 and $\sim$0.18 mJy beam$^{-1}$ at 54 and 144 MHz, respectively. The yellow circles denote 1 $r_\mathrm{e}$ for both the N and the S halos.}
\label{fig:PSZ107}
\end{figure}

\begin{figure}[h!]
\centering
{\includegraphics[scale=0.18]{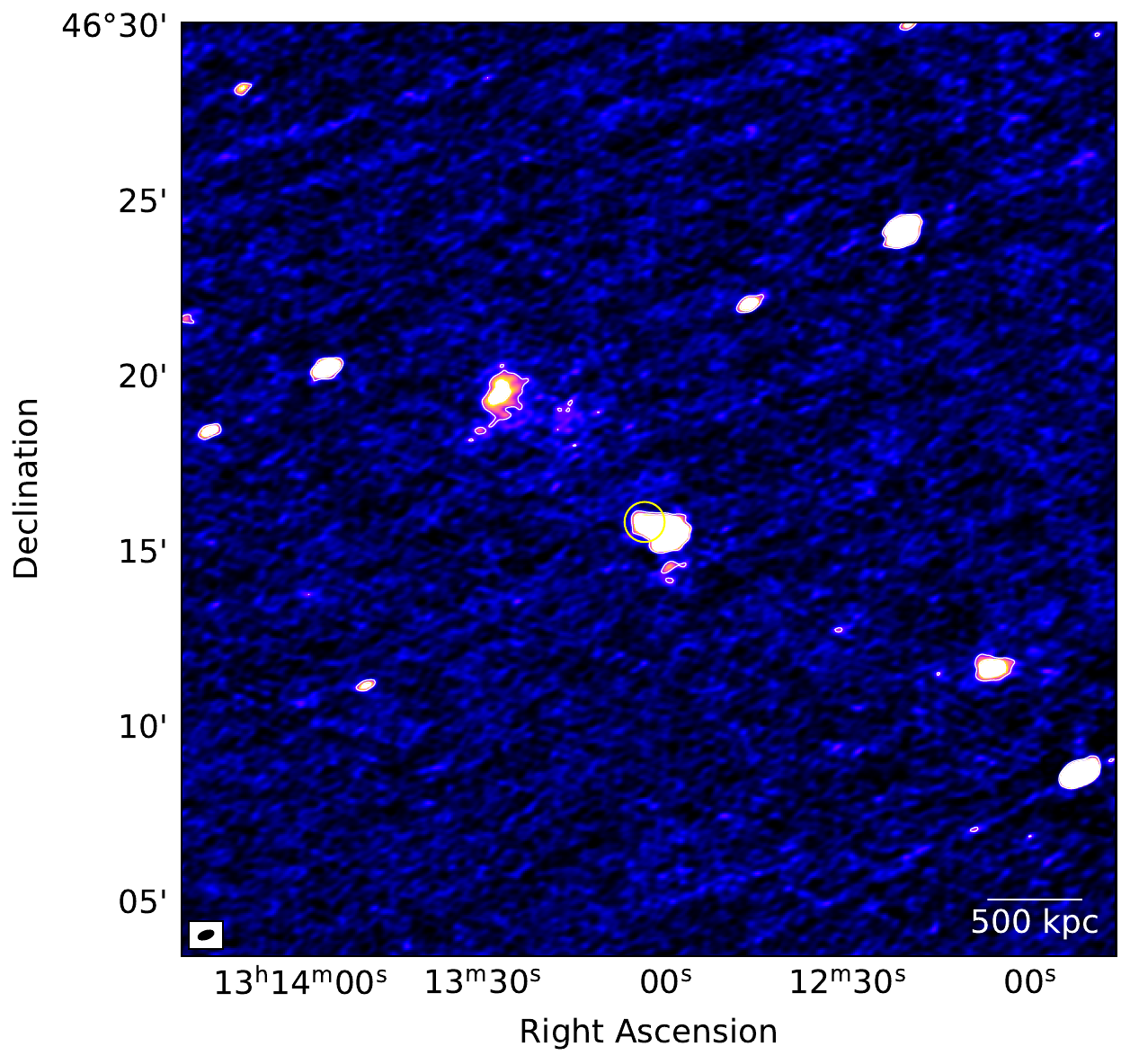}}
{\includegraphics[scale=0.18]{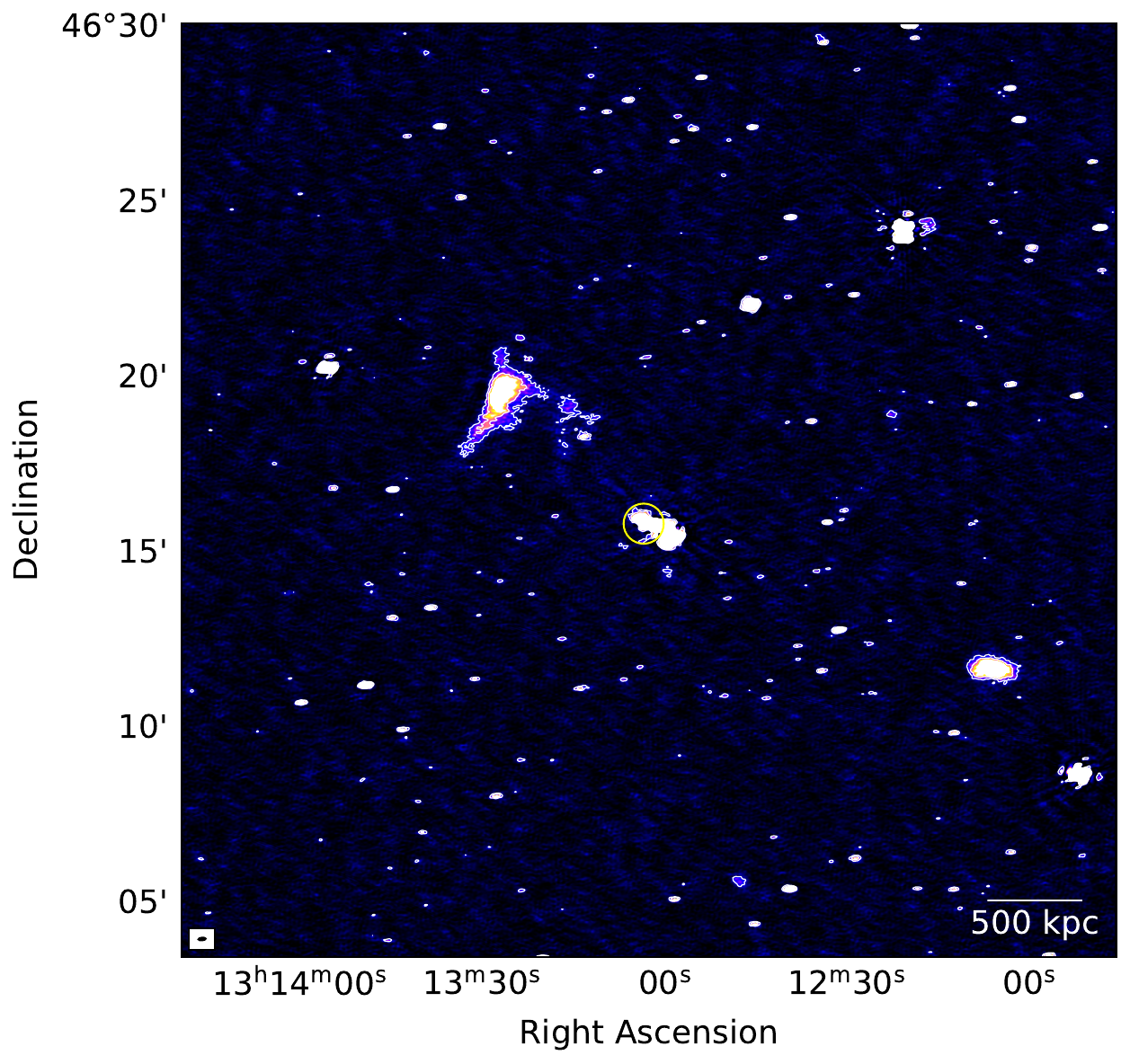}}
{\includegraphics[scale=0.18]{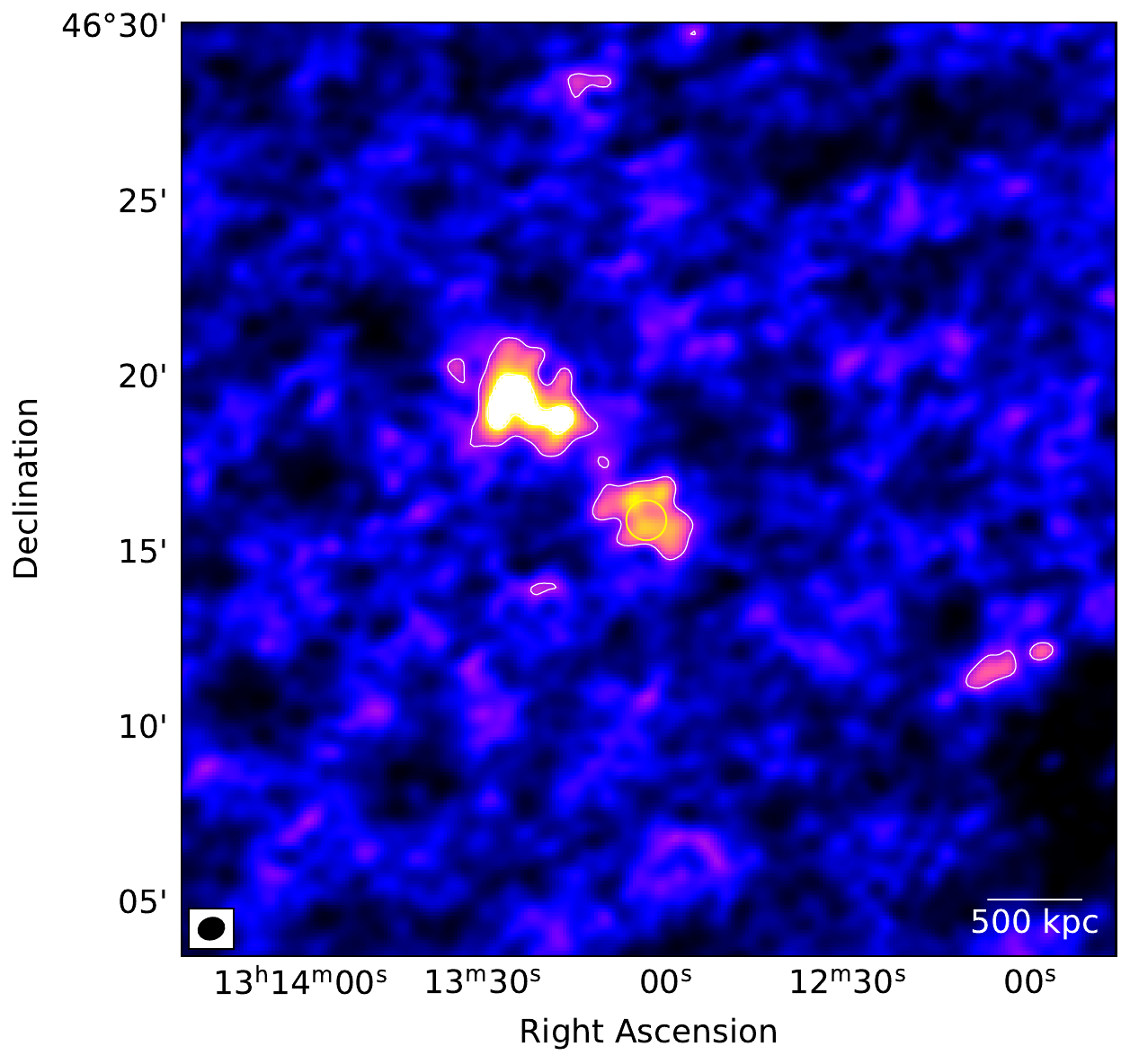}}
{\includegraphics[scale=0.18]{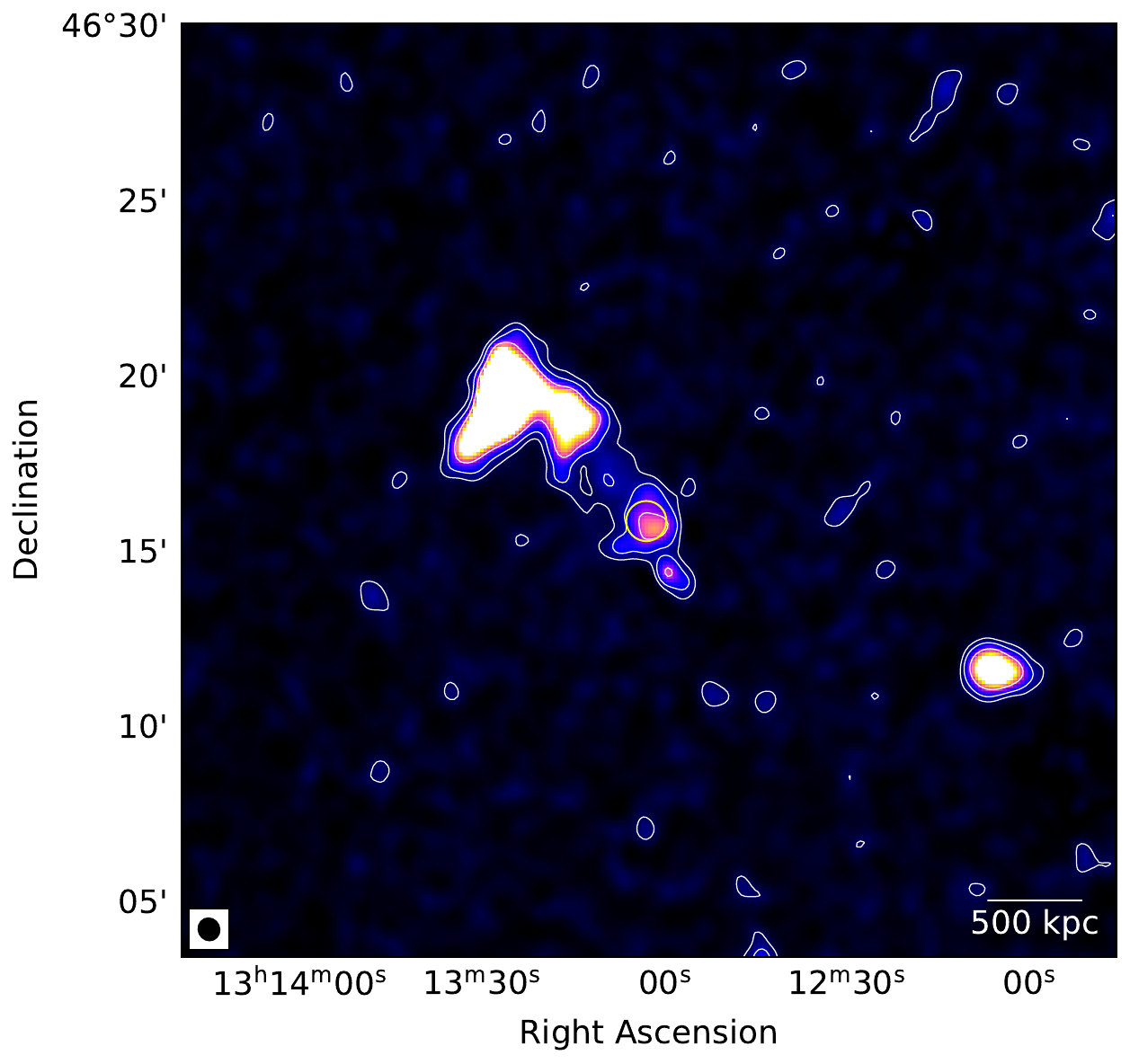}}
\caption{\footnotesize \textit{From left to right}: 54 MHz briggs, 144 MHz briggs, 54 MHz low-resolution and 144 MHz low-resolution images of PSZ2G111.75+70.37. Low-resolution images were produced by tapering visibilities at an angular scale corresponding to 100 kpc at the cluster redshift. Their \textit{rms} noise is $\sim$4.4 and $\sim$0.22 mJy beam$^{-1}$ at 54 and 144 MHz, respectively. The yellow circle denotes 1 $r_\mathrm{e}$.}
\label{fig:PSZ111}
\end{figure}

\begin{figure}[h!]
\centering
{\includegraphics[scale=0.18]{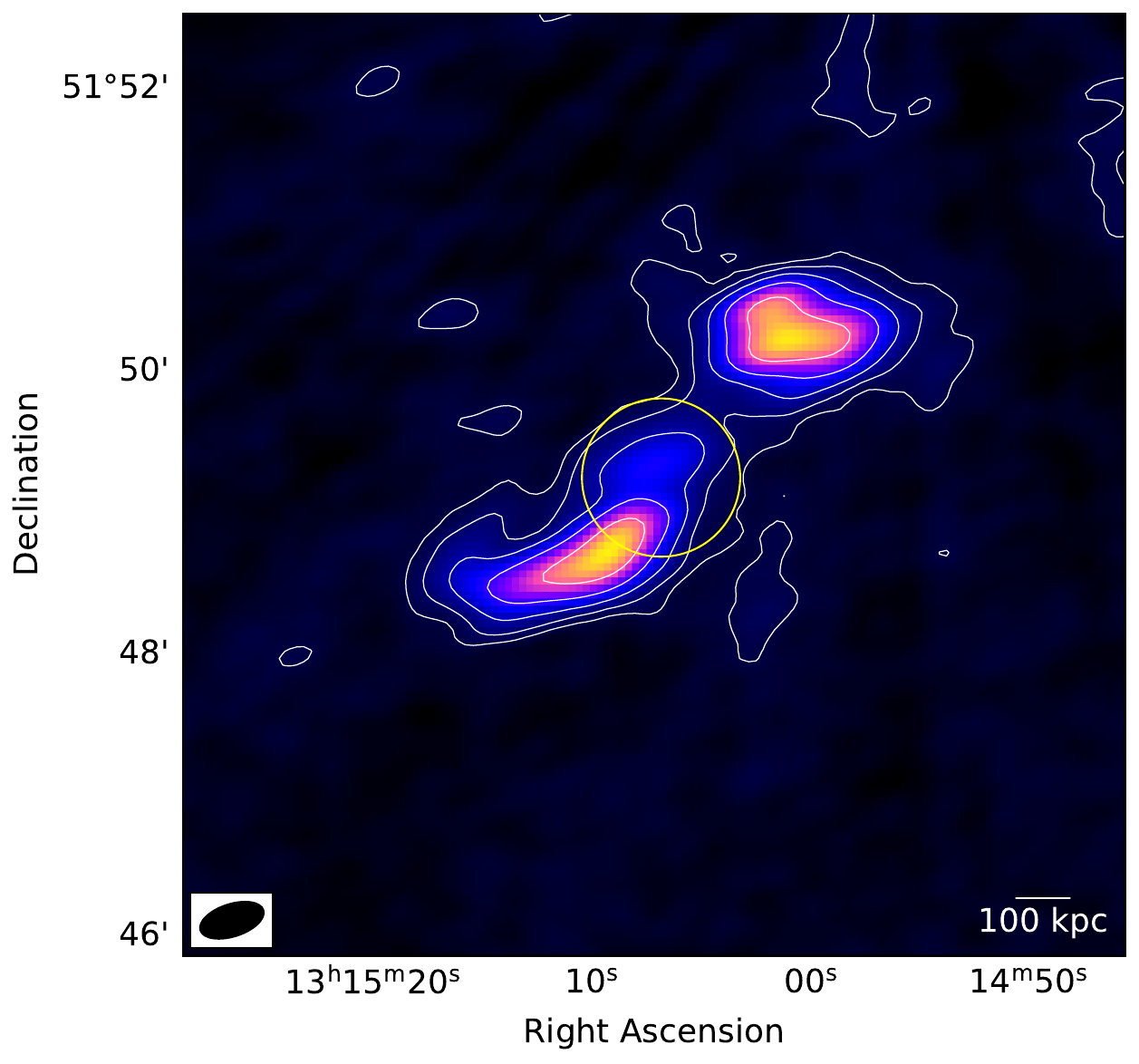}}
{\includegraphics[scale=0.18]{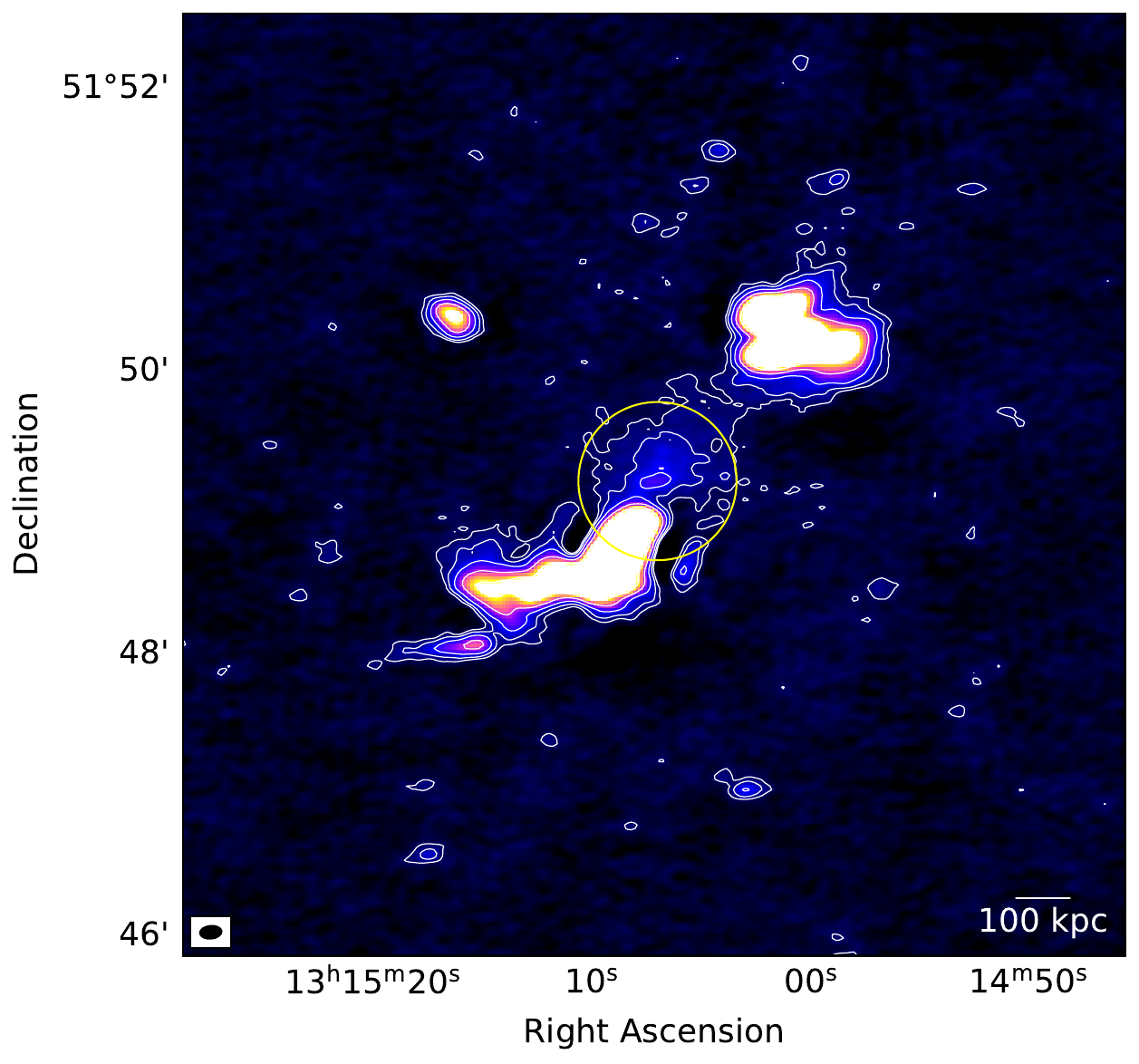}}
{\includegraphics[scale=0.18]{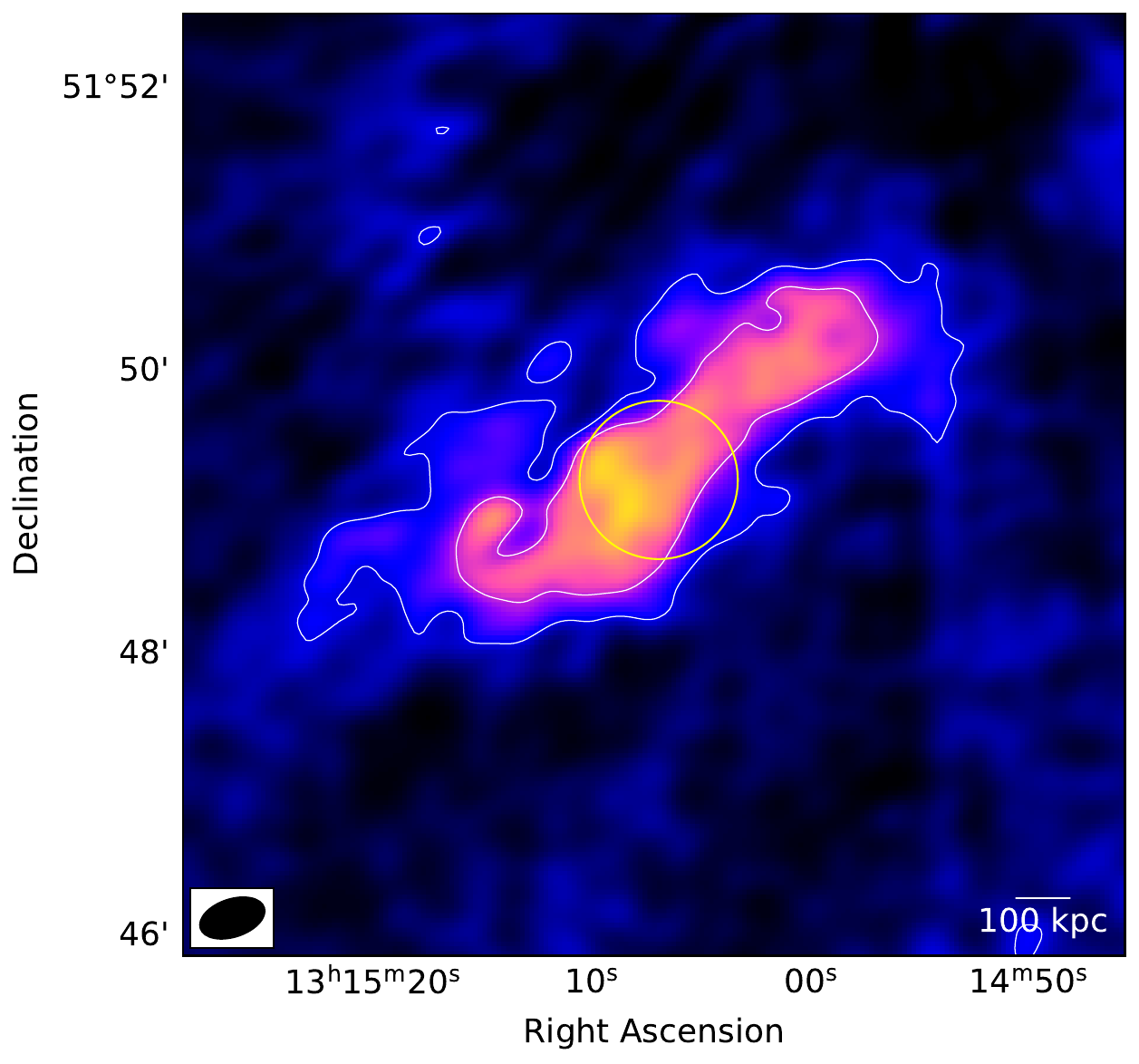}}
{\includegraphics[scale=0.18]{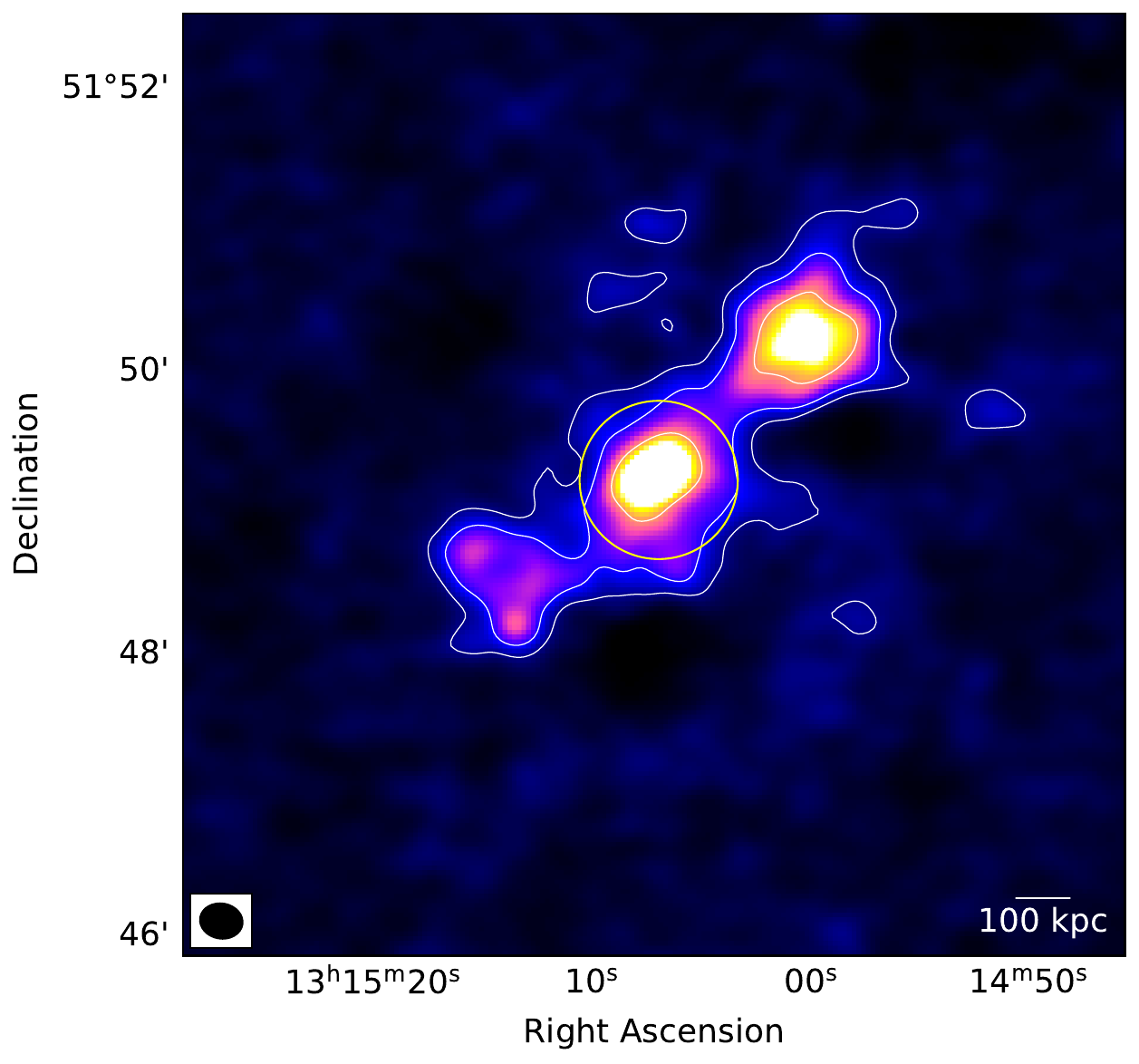}}
\caption{\footnotesize \textit{From left to right}: 54 MHz briggs, 144 MHz briggs, 54 MHz low-resolution and 144 MHz low-resolution images of PSZ2G114.31+64.89. Low-resolution images were produced by tapering visibilities at an angular scale corresponding to 50 kpc at the cluster redshift. Their \textit{rms} noise is $\sim$2.4 and $\sim$0.1 mJy beam$^{-1}$ at 54 and 144 MHz, respectively. The yellow circle denotes 1 $r_\mathrm{e}$.}
\label{fig:PSZ114}
\end{figure}

\newpage
\begin{figure}[h!]
\centering
{\includegraphics[scale=0.18]{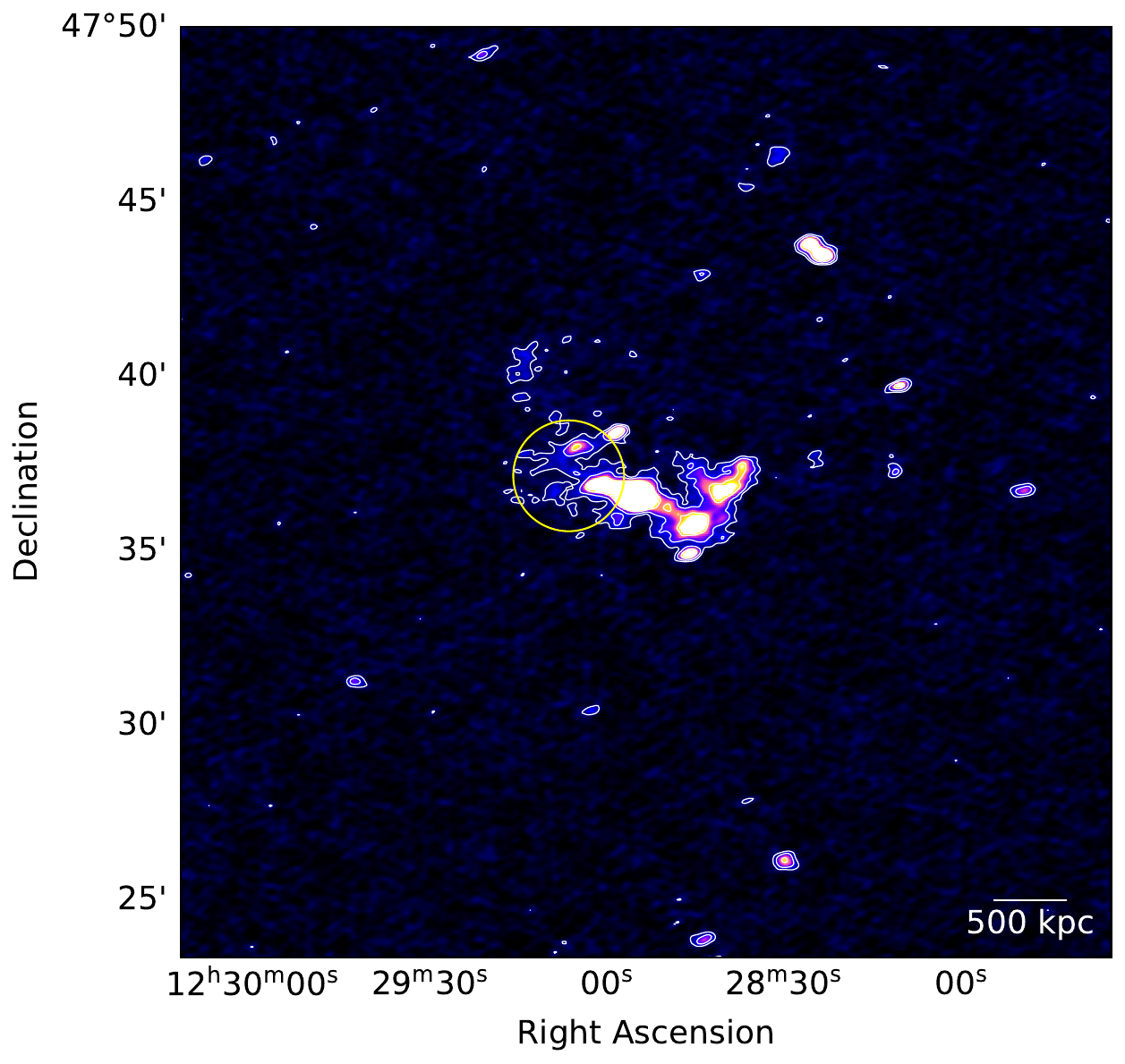}}
{\includegraphics[scale=0.18]{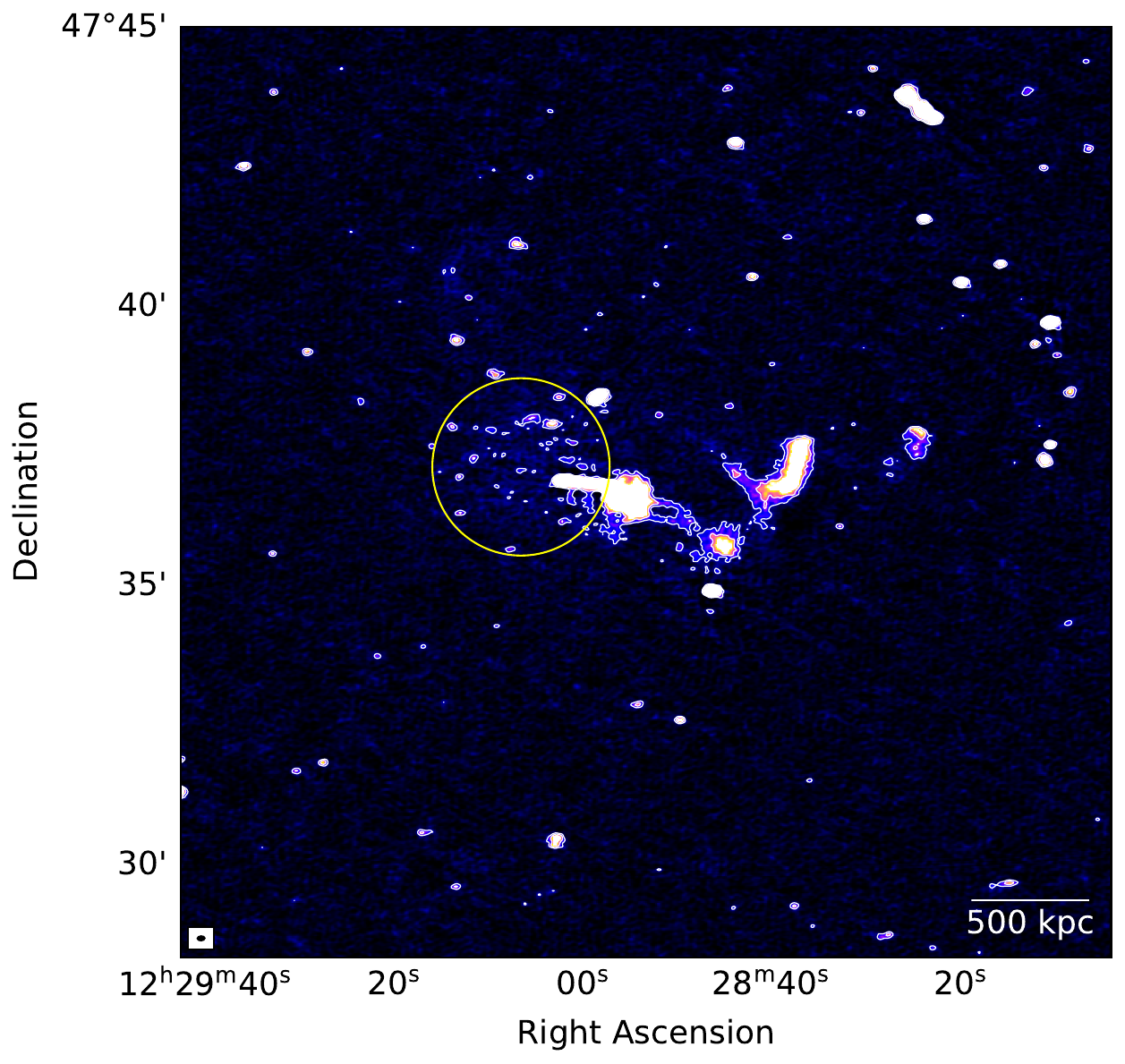}}
{\includegraphics[scale=0.18]{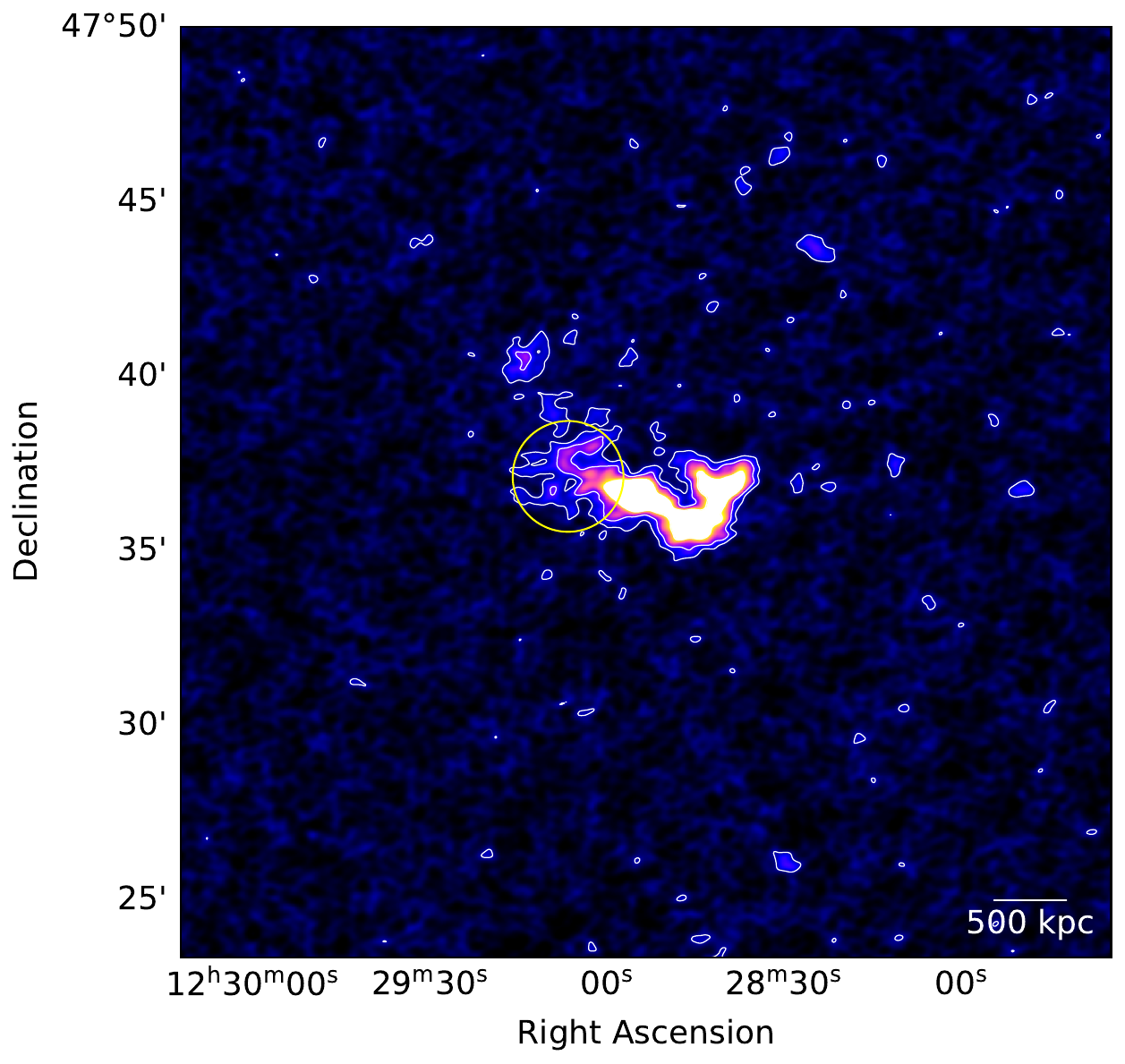}}
{\includegraphics[scale=0.18]{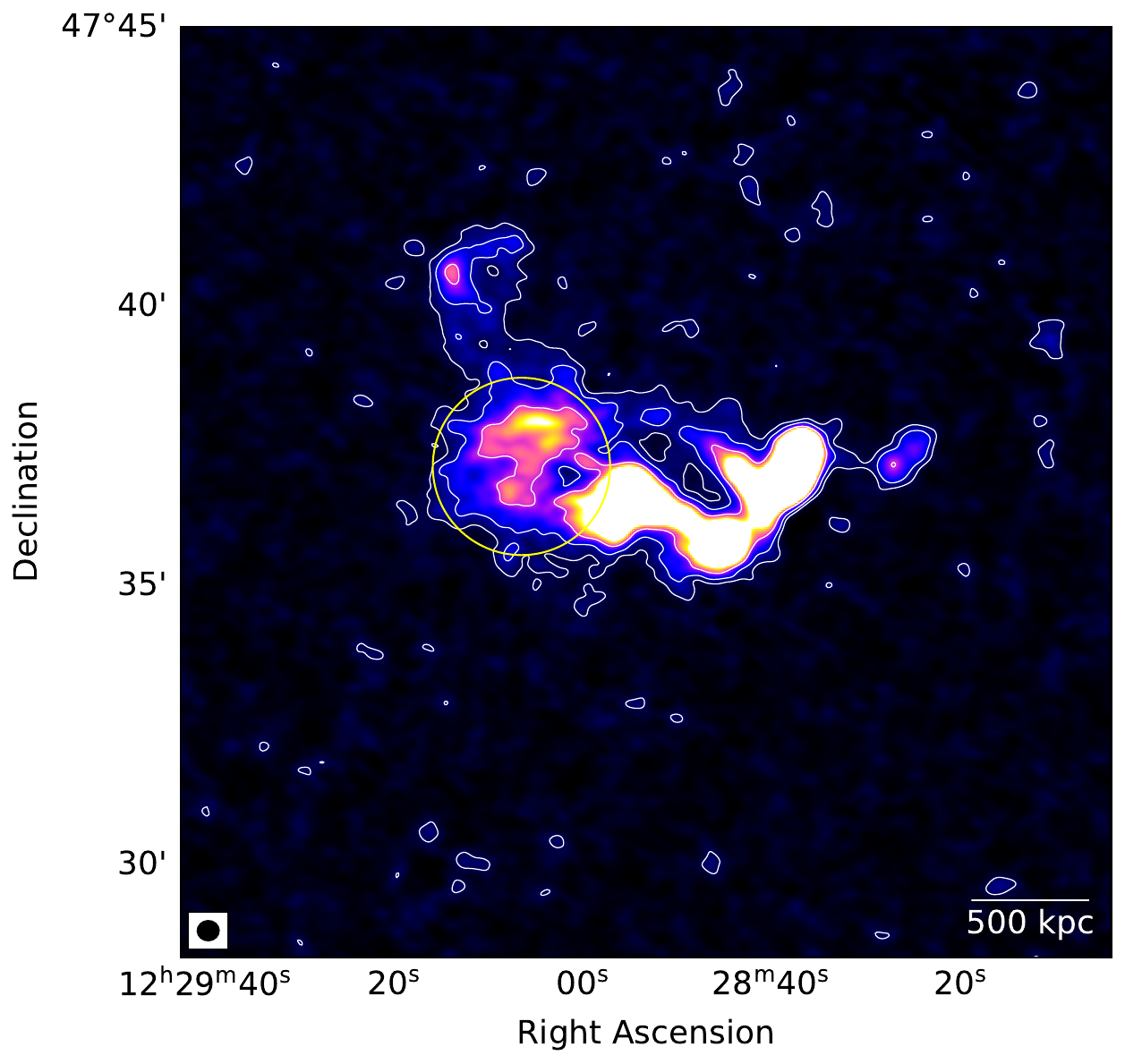}}
\caption{\footnotesize \textit{From left to right}: 54 MHz briggs, 144 MHz briggs, 54 MHz low-resolution and 144 MHz low-resolution images of PSZ2G133.60+69.04. Low-resolution images were produced by tapering visibilities at an angular scale corresponding to 50 kpc at the cluster redshift. Their \textit{rms} noise is $\sim$1.5 and $\sim$0.12 mJy beam$^{-1}$ at 54 and 144 MHz, respectively. The yellow circle denotes 1 $r_\mathrm{e}$.}
\label{fig:PSZ133}
\end{figure}

\begin{figure}[h!]
\centering
{\includegraphics[scale=0.18]{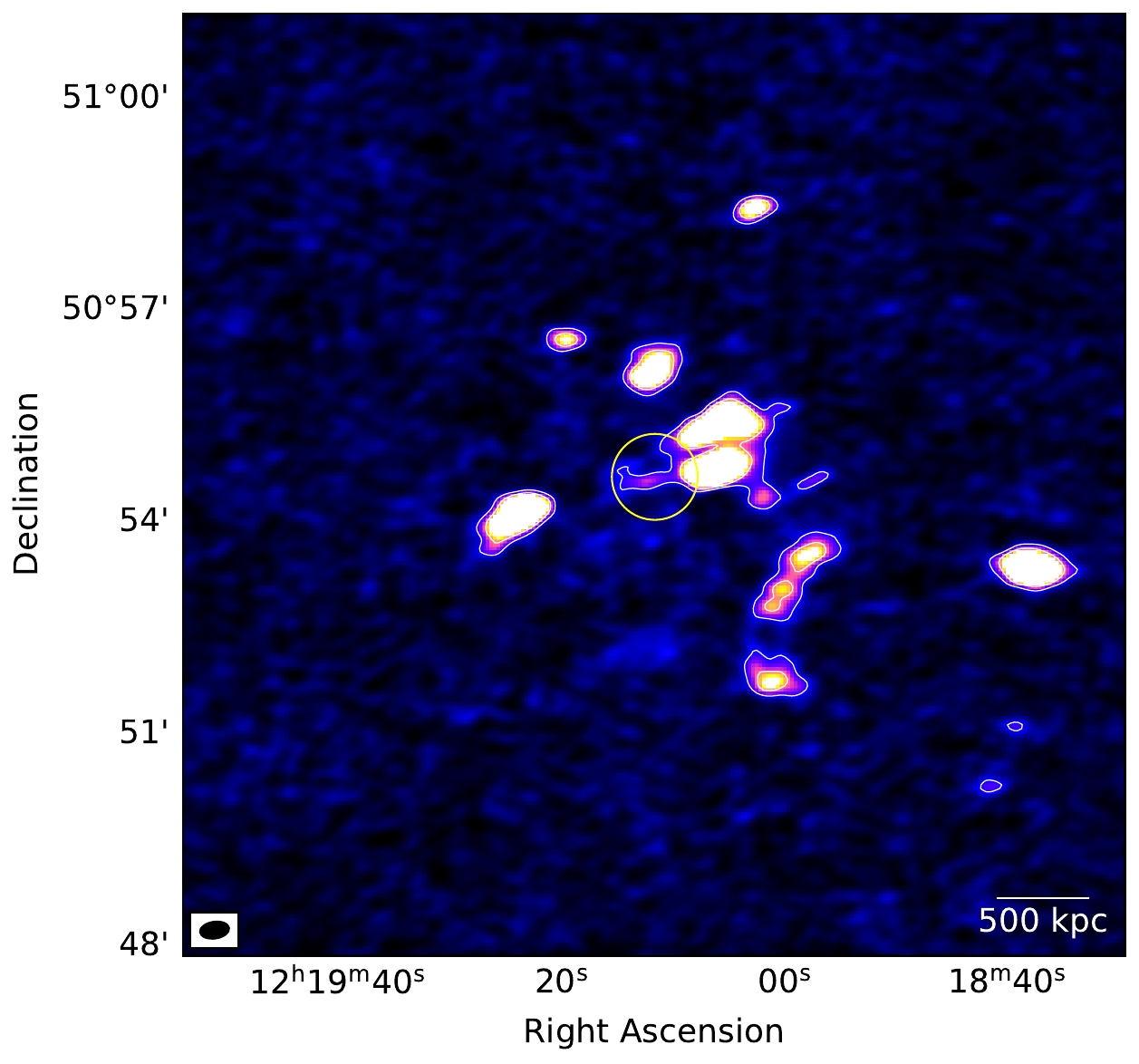}}
{\includegraphics[scale=0.18]{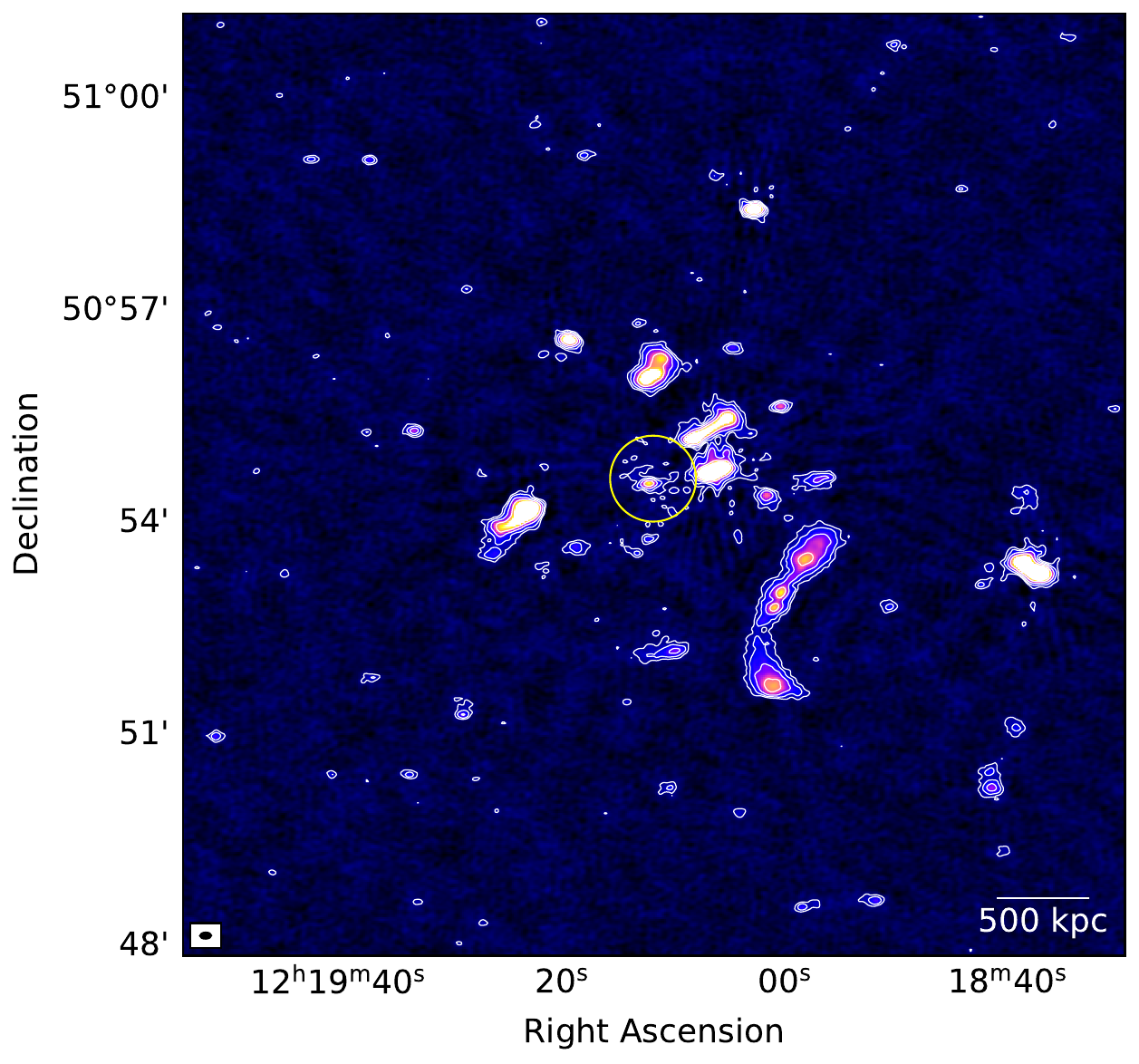}}
{\includegraphics[scale=0.18]{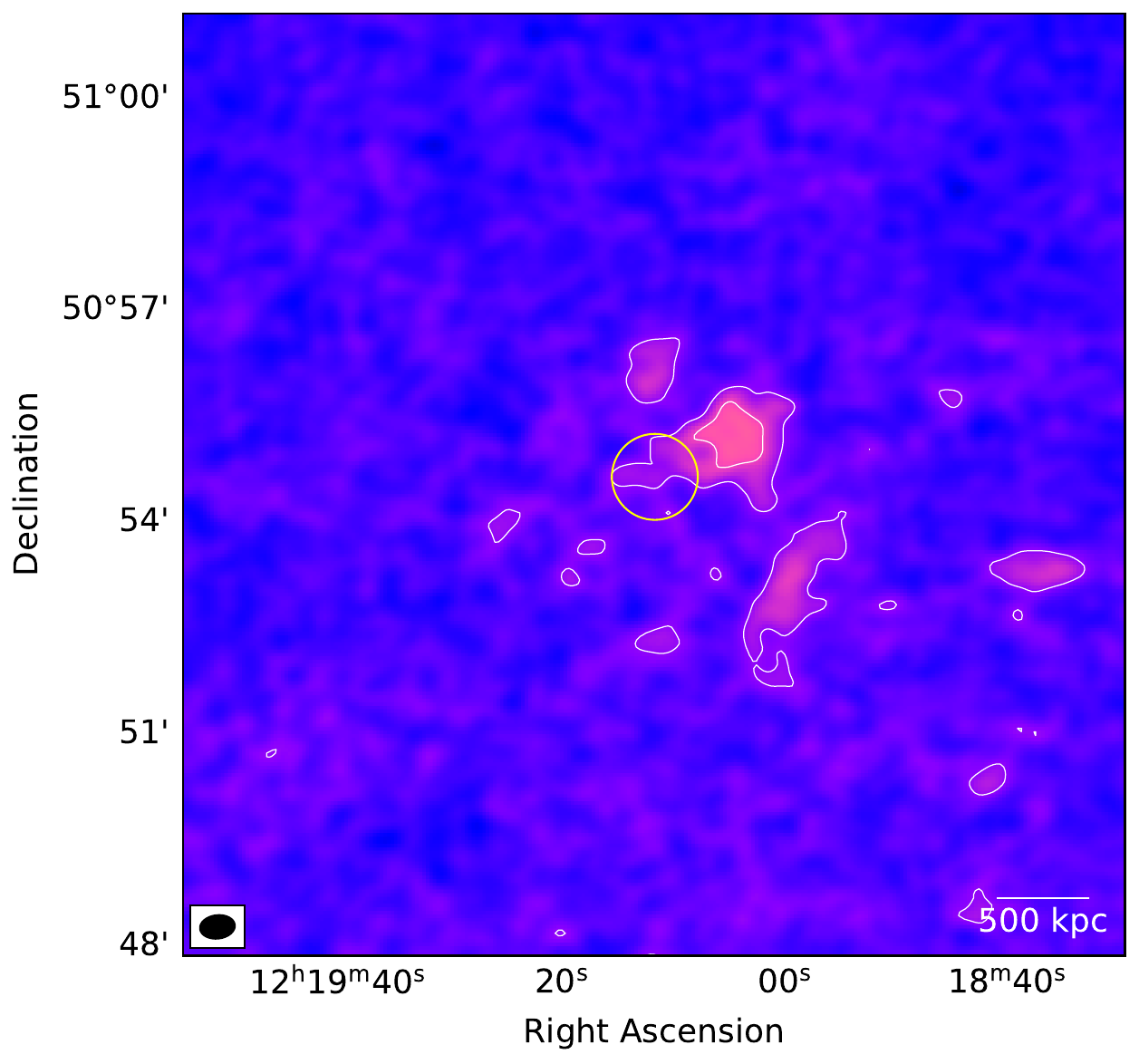}}
{\includegraphics[scale=0.18]{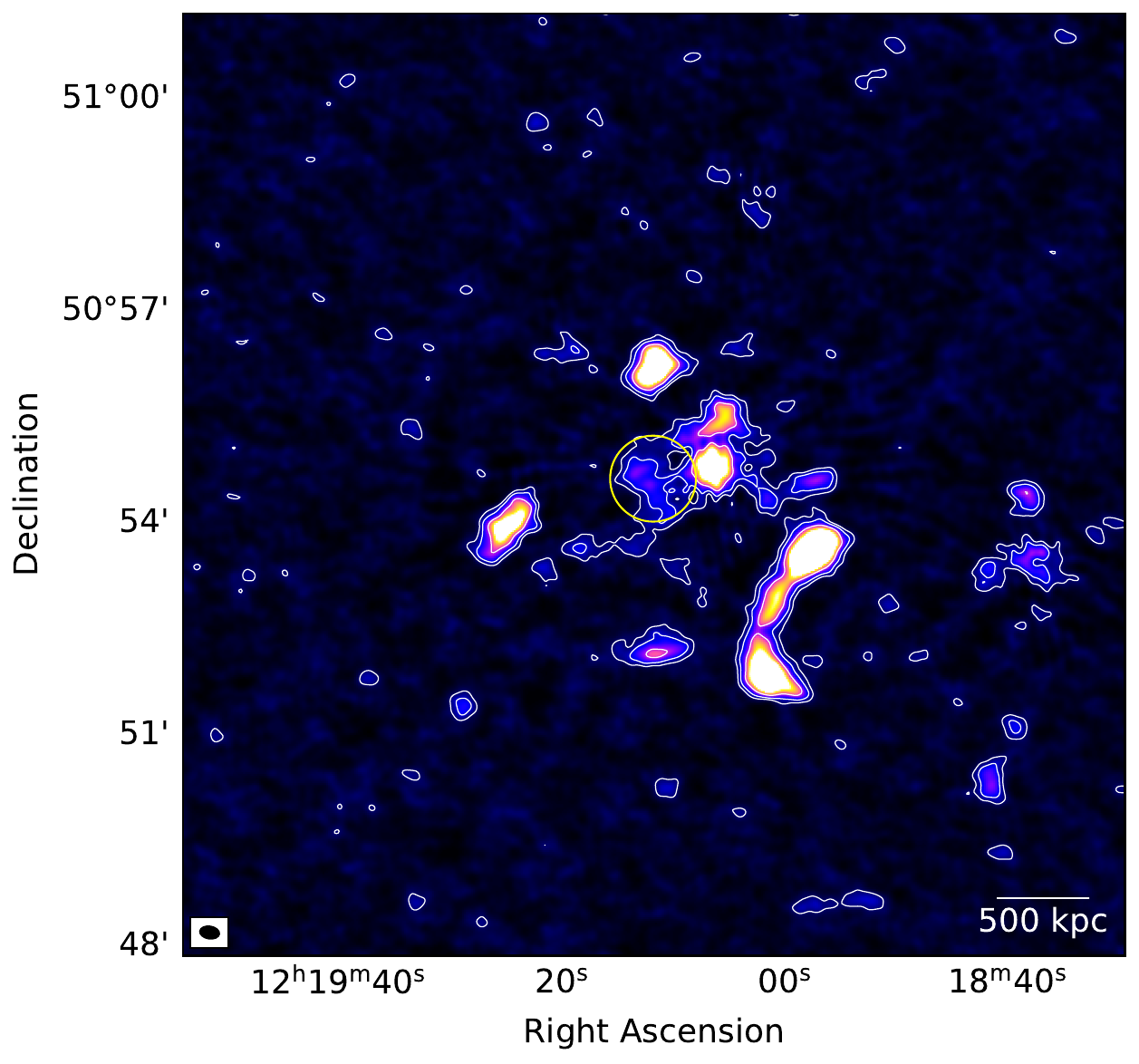}}
\caption{\footnotesize \textit{From left to right}: 54 MHz briggs, 144 MHz briggs, 54 MHz low-resolution and 144 MHz low-resolution images of PSZ2G135.17+65.43. Low-resolution images were produced by tapering visibilities at an angular scale corresponding to 50 kpc at the cluster redshift. Their \textit{rms} noise is $\sim$1.5 and $\sim$0.11 mJy beam$^{-1}$ at 54 and 144 MHz, respectively. The yellow circle denotes 1 $r_\mathrm{e}$.}
\label{fig:PSZ135}
\end{figure}

\begin{figure}[h!]
\centering
{\includegraphics[scale=0.18]{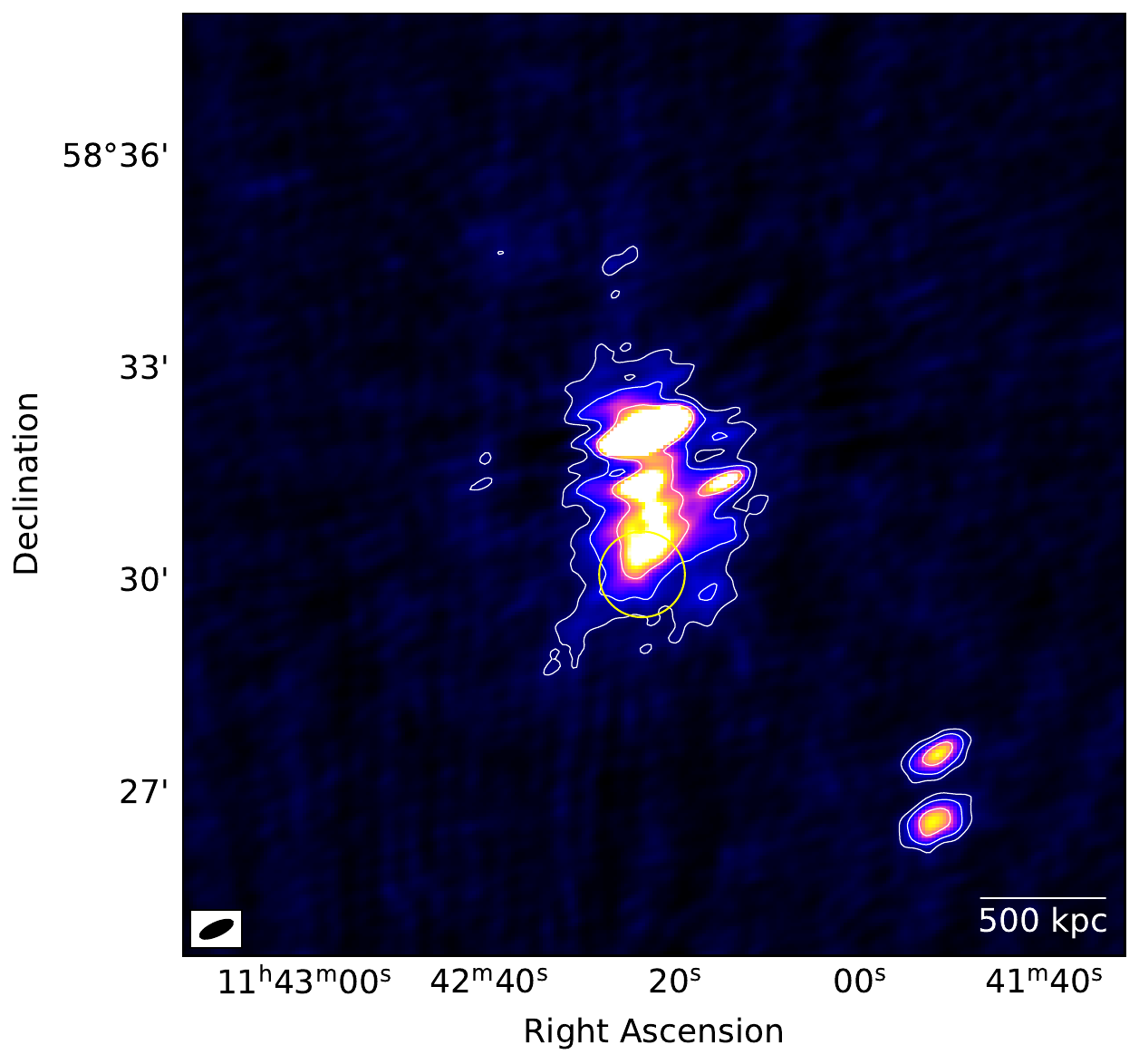}}
{\includegraphics[scale=0.18]{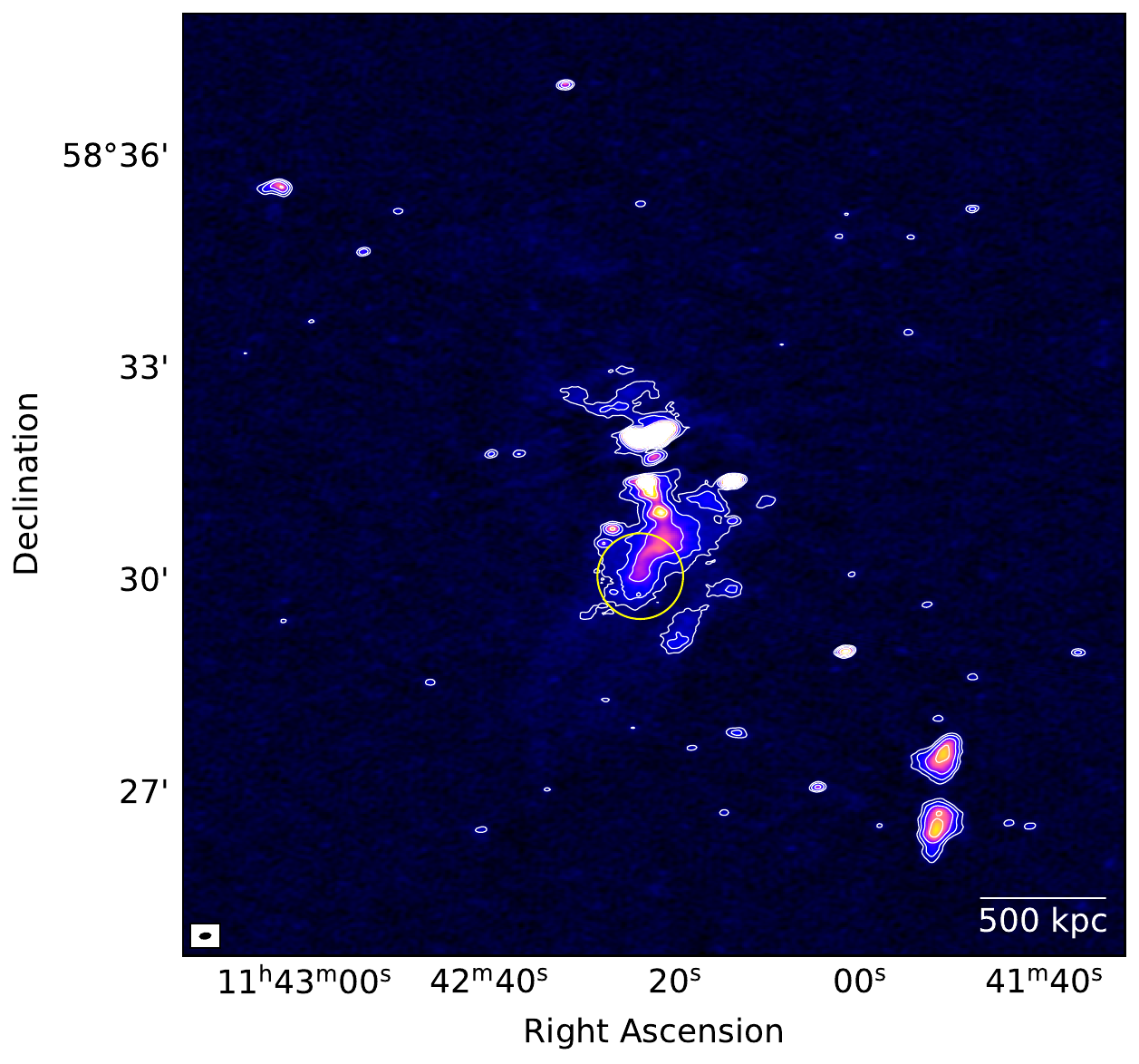}}
{\includegraphics[scale=0.18]{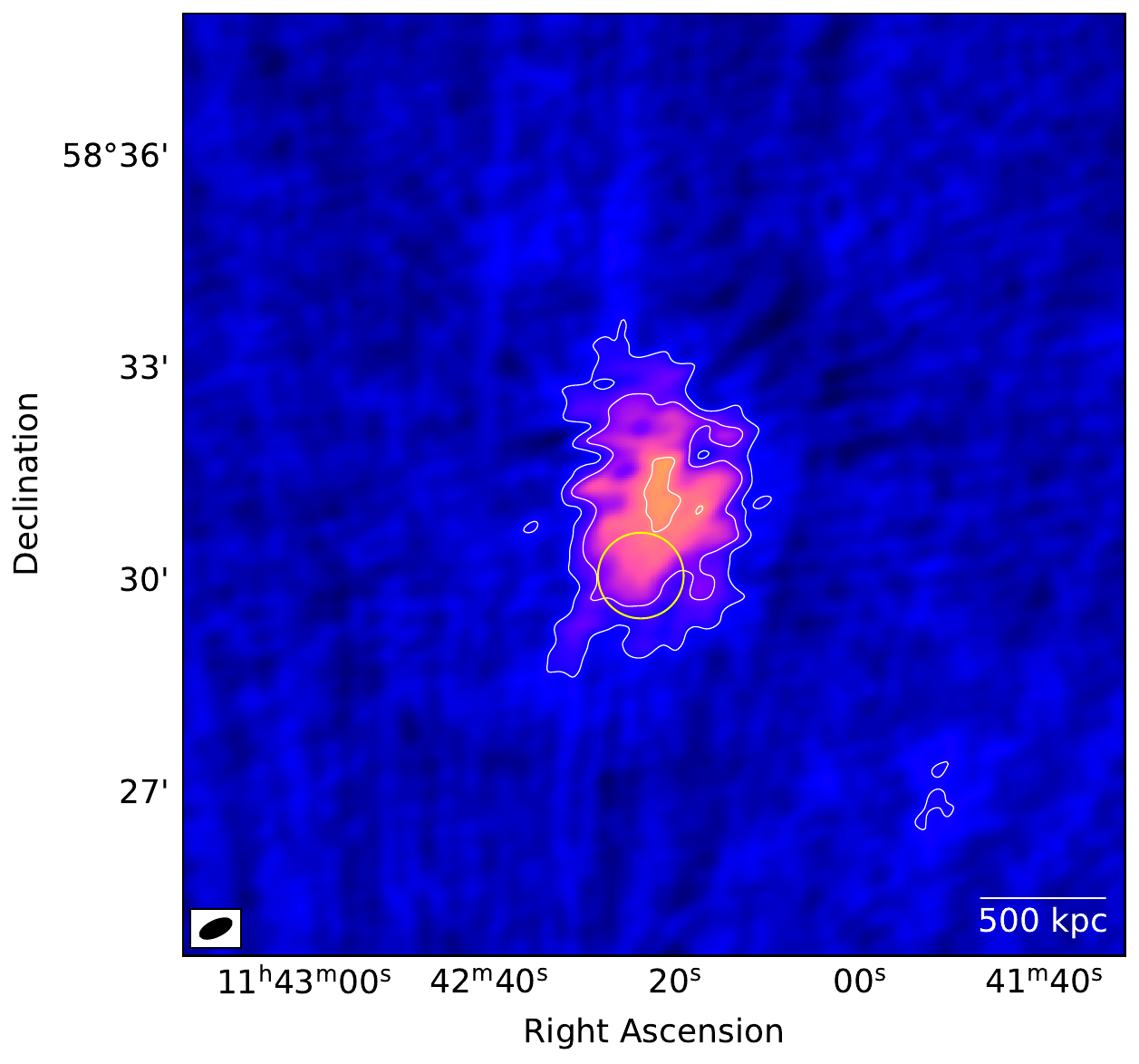}}
{\includegraphics[scale=0.18]{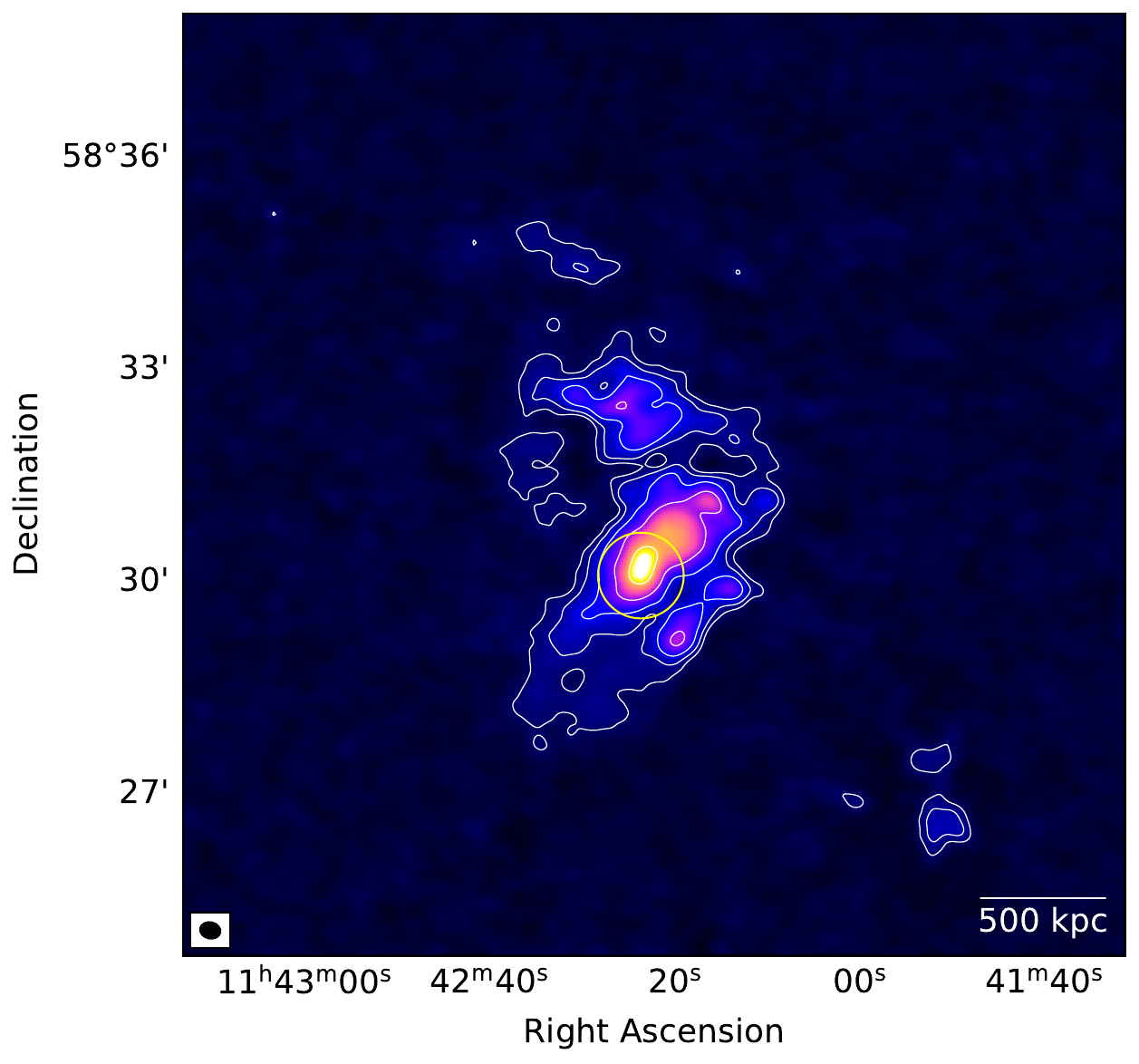}}
\caption{\footnotesize \textit{From left to right}: 54 MHz briggs, 144 MHz briggs, 54 MHz low-resolution and 144 MHz low-resolution images of PSZ2G139.18+56.37. Low-resolution images were produced by tapering visibilities at an angular scale corresponding to 50 kpc at the cluster redshift. Their \textit{rms} noise is $\sim$2.3 and $\sim$0.09 mJy beam$^{-1}$ at 54 and 144 MHz, respectively. The yellow circle denotes 1 $r_\mathrm{e}$.}
\label{fig:PSZ139}
\end{figure}

\begin{figure}[h!]
\centering
{\includegraphics[scale=0.18]{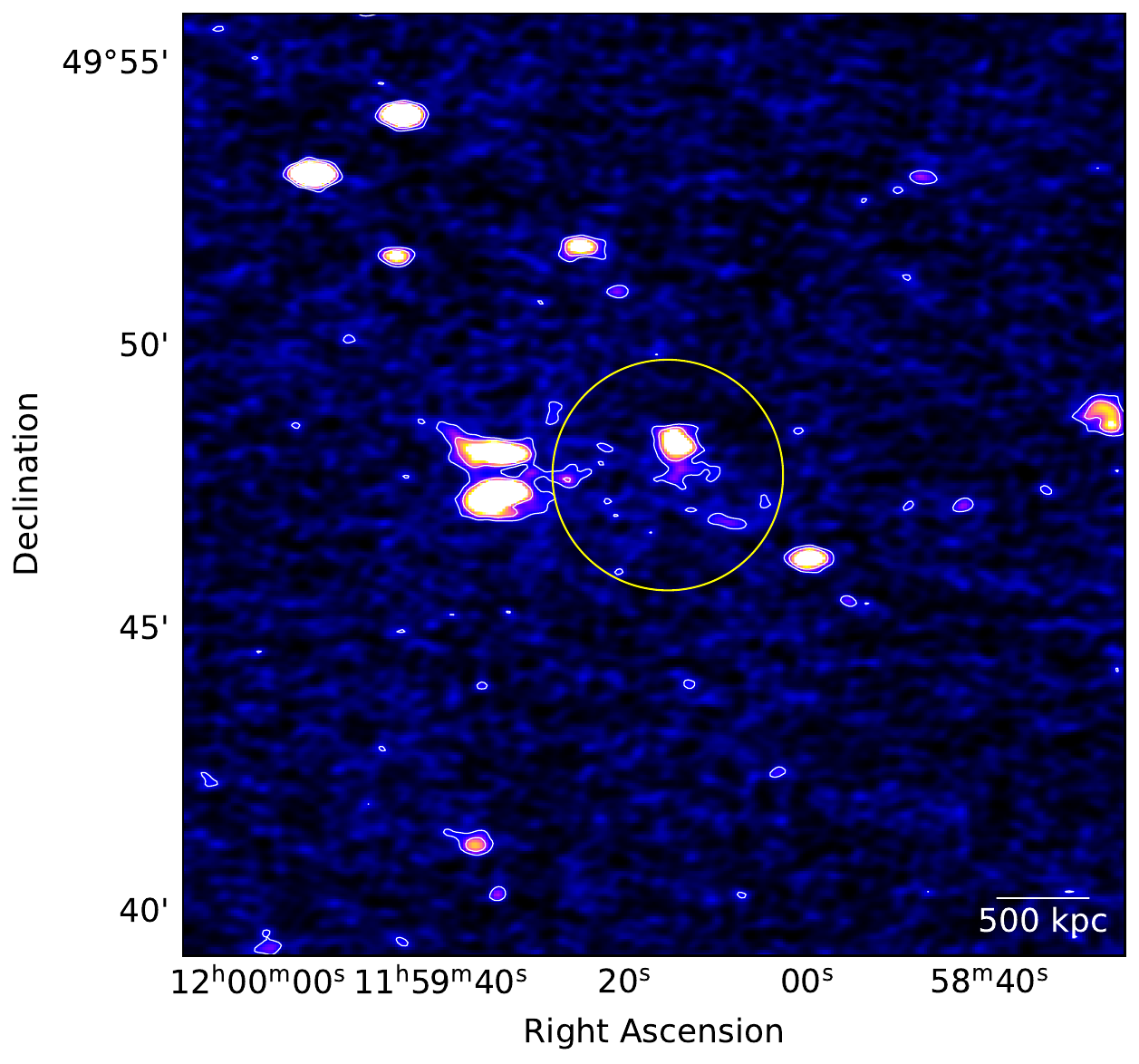}}
{\includegraphics[scale=0.18]{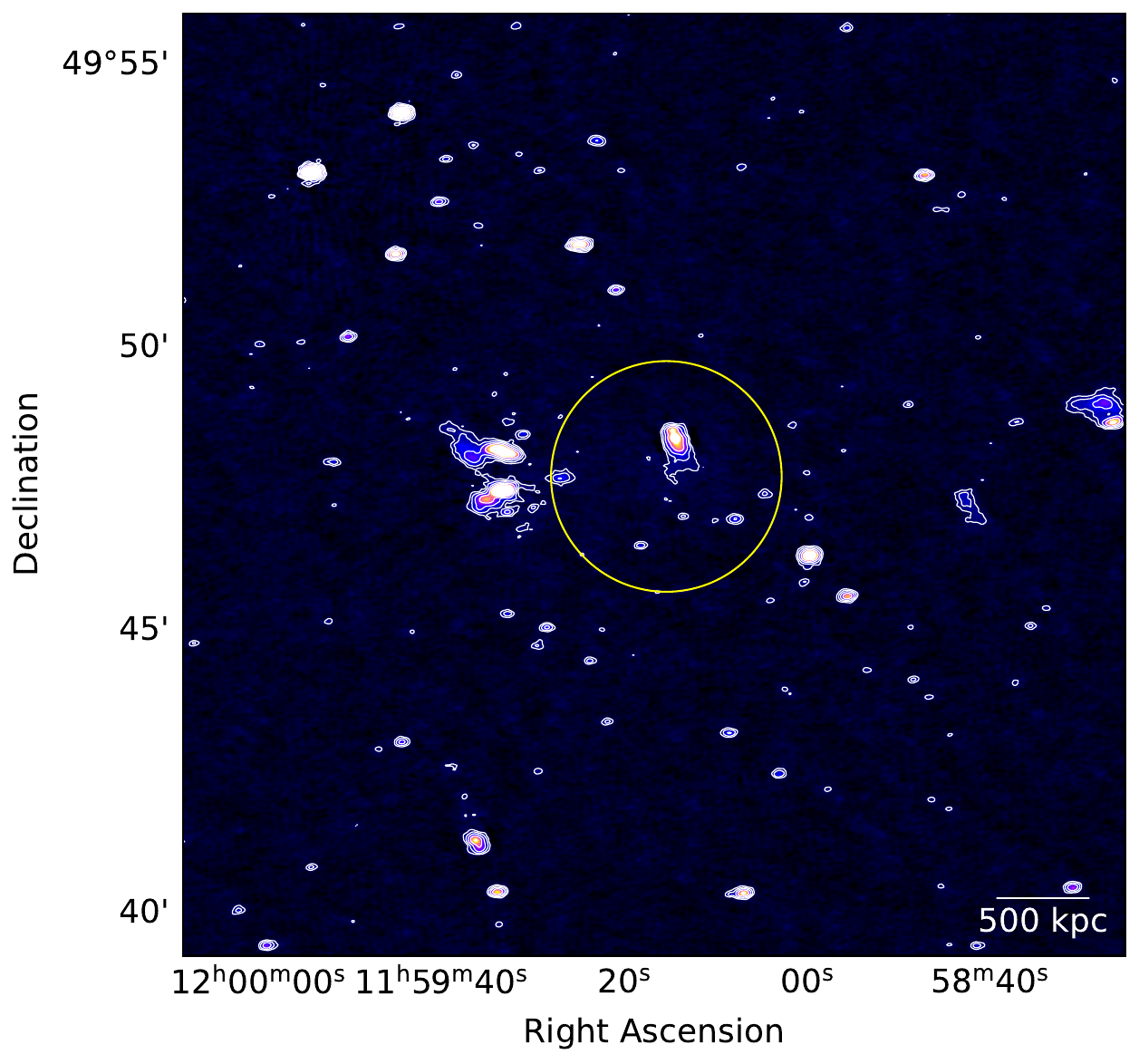}}
{\includegraphics[scale=0.18]{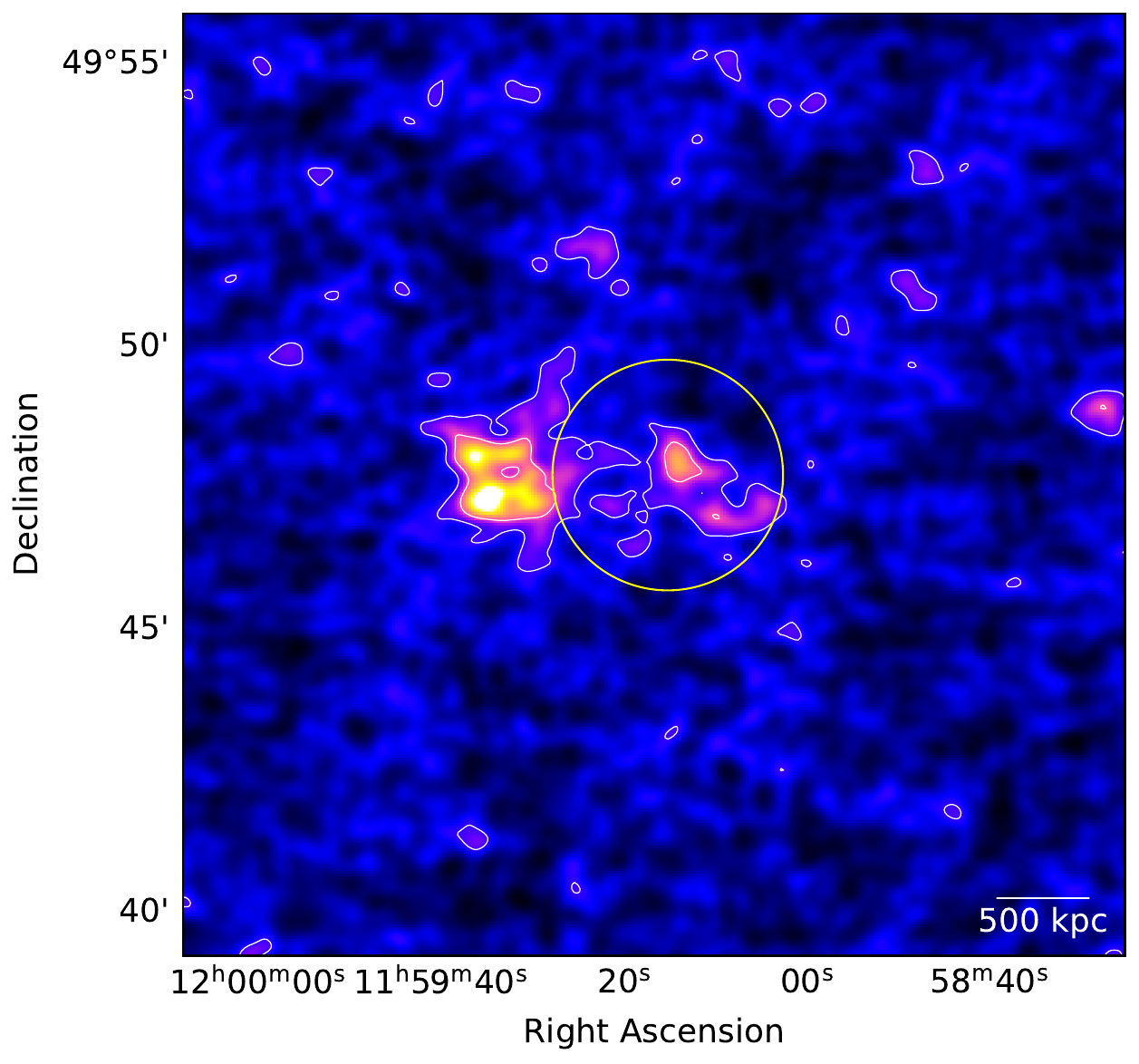}}
{\includegraphics[scale=0.18]{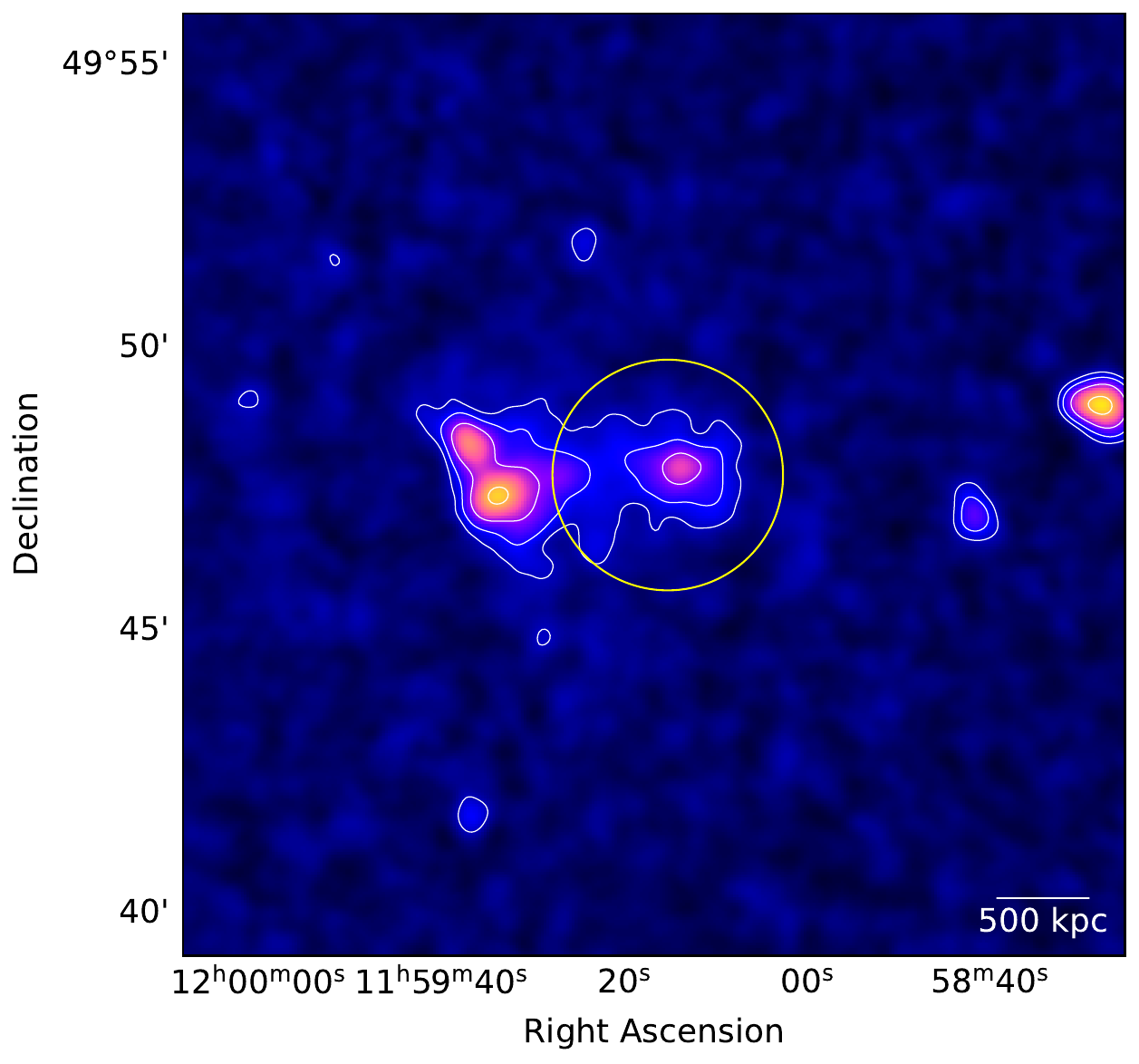}}
\caption{\footnotesize \textit{From left to right}: 54 MHz briggs, 144 MHz briggs, 54 MHz low-resolution and 144 MHz low-resolution images of PSZ2G143.26+65.24. Low-resolution images were produced by tapering visibilities at an angular scale corresponding to 100 kpc at the cluster redshift. Their \textit{rms} noise is $\sim$1.8 and $\sim$0.17 mJy beam$^{-1}$ at 54 and 144 MHz, respectively. The yellow circle denotes 1 $r_\mathrm{e}$.}
\label{fig:PSZ143}
\end{figure}

\end{appendix}

\end{document}